\documentclass[12pt,indent]{article}

\usepackage{a4,latexsym,amsmath,amssymb}
\usepackage[latin1]{inputenc}
\usepackage{float}
\usepackage{natbib}
\usepackage{graphics}
\usepackage[english]{babel}
\usepackage{tabularx}
\usepackage{lscape}
\usepackage{multirow}
\usepackage{rotating}
\usepackage{dsfont}
\usepackage{subfig}
\usepackage{bbm}
\usepackage{booktabs}
\usepackage[table]{xcolor}

\usepackage{calc}
\usepackage{ifthen}
\newenvironment{tabularsmall}
{ \footnotesize \sffamily \tabular } {
\endtabular
\normalfont }

\newcommand{\alphab}{\boldsymbol{\alpha}}

\newcommand{\deltab}{\boldsymbol{\delta}}

\newcommand{\gammab}{\boldsymbol{\gamma}}

\newcommand{\xb}{\boldsymbol{x}}

\newcommand{\0}{\textbf{0}}

\newcommand{\blanco}[1]{}

\def\d{\displaystyle}

\usepackage{natbib}

\usepackage{setspace}

\begin{document}
\bibliographystyle{chicago}
\sloppy

\makeatletter
\renewcommand{\section}{\@startsection{section}{1}{\z@}%
        {-3.5ex \@plus -1ex \@minus -.2ex}%
        {1.5ex \@plus.2ex}%
        {\reset@font\Large\sffamily}}
\renewcommand{\subsection}{\@startsection{subsection}{1}{\z@}%
        {-3.25ex \@plus -1ex \@minus -.2ex}%
        {1.1ex \@plus.2ex}%
        {\reset@font\large\sffamily\flushleft}}
\renewcommand{\subsubsection}{\@startsection{subsubsection}{1}{\z@}%
        {-3.25ex \@plus -1ex \@minus -.2ex}%
        {1.1ex \@plus.2ex}%
        {\reset@font\normalsize\sffamily\flushleft}}
\makeatother



\newsavebox{\tempbox}
\newlength{\linelength}
\setlength{\linelength}{\linewidth-10mm} \makeatletter
\renewcommand{\@makecaption}[2]
{
  \renewcommand{\baselinestretch}{1.1} \normalsize\small
  \vspace{5mm}
  \sbox{\tempbox}{#1: #2}
  \ifthenelse{\lengthtest{\wd\tempbox>\linelength}}
  {\noindent\hspace*{4mm}\parbox{\linewidth-10mm}{\sc#1: \sl#2\par}}
  {\begin{center}\sc#1: \sl#2\par\end{center}}
}



\def\R{\mathchoice{ \hbox{${\rm I}\!{\rm R}$} }
                   { \hbox{${\rm I}\!{\rm R}$} }
                   { \hbox{$ \scriptstyle  {\rm I}\!{\rm R}$} }
                   { \hbox{$ \scriptscriptstyle  {\rm I}\!{\rm R}$} }  }

\def\N{\mathchoice{ \hbox{${\rm I}\!{\rm N}$} }
                   { \hbox{${\rm I}\!{\rm N}$} }
                   { \hbox{$ \scriptstyle  {\rm I}\!{\rm N}$} }
                   { \hbox{$ \scriptscriptstyle  {\rm I}\!{\rm N}$} }  }

\def\d{\displaystyle}

\title{Detection of  Uniform and Non-Uniform Differential Item Functioning by Item Focussed Trees
}
\author{ Moritz Berger  \& Gerhard Tutz\\{\small Ludwig-Maximilians-Universit\"{a}t M\"{u}nchen}\\
{\small Akademiestra{\ss}e 1, 80799 M\"{u}nchen}}

\maketitle

\begin{abstract}
\noindent
Detection of differential item functioning by use of the logistic modelling approach has a long tradition.
One big advantage of the approach is that it can be used to investigate non-uniform DIF as well as uniform DIF. The classical approach allows to detect DIF by distinguishing between multiple groups. We propose an alternative method that is a combination  of recursive partitioning methods (or trees) and logistic regression methodology to detect uniform and non-uniform DIF in a nonparametric way. The output of the method are trees that visualize in a simple way the structure of DIF in an item showing which variables are interacting in which way when generating DIF. In addition we consider a logistic regression method in which DIF can by induced by a vector of covariates, which may include categorical but also continuous covariates. The methods are investigated in simulation studies and illustrated by two applications.
\end{abstract}

\noindent{\bf Keywords:} Logistic regression; Differential item functioning; Recursive partitioning; Item focussed Trees

\section{Introduction}\label{sec:introduction}

In recent years differential item functioning (DIF) and DIF identification methods have been areas of intensive current research.  Differential item functioning occurs if
the probability of a correct response among
persons with the same value of their underlying trait differs in subgroups, for example, if the difficulty of an item
depends on the membership to a  racial, ethnic or gender
subgroup.   The effect is measurement bias and possibly
discrimination, see, for example, \citet{wainer1993differential,osterlind2009differential,rogers2005differential},
\citet{millsap1993methodology}, \citet{zumbo1999handbook}.

A variety of methods to detect DIF has been proposed, for a more recent overview see \citet{magisetal:2010}. One can in particular distinguish between \textit{item response theory (IRT) modelling approaches} and \textit{test score methods}   \citep{magis2014detection}.
The former assume that an IRT model holds in each group. Tests as Lord`s test or likelihood ratio tests are used to detect differences of item parameters between groups. IRT approaches have been used, among others, by  \citet{Lord:1980}, \citet{raju1988area} and \citet{wainer1993differential}. Test score methods use a matching variable as, for example, Mantel-Haenzel test procedures \citep{holland1988} or logistic regression modelling \citep{swaminathan1990detecting}.  We will use the logistic regression framework since it also allows to investigate non-uniform DIF.
Uniform DIF is present if the (scaled) differences  in the probabilities of solving an item of subjects from different groups  but with the same ability level do not depend on the common ability level. In non-uniform DIF scenarios the differences are not constant across ability levels and crossing item response curves may occur.

More recently IRT based DIF modelling has been extended to allow for continuous variables that induce DIF. The corresponding latent trait models contain many parameters since each item comes with an own vector of parameters. Therefore maximum likelihood estimates are bound to fail. \citet{TuSchauDIFPsych} used a penalty approach to regularize parameter estimation, \citet{SchauTuBoost2015} used boosting techniques whereas \citet{TuBerg2015IF} rely on recursive partitioning methods. A non-IRT modelling approach with regularization by penalties has been proposed by \citet{magis2014detection}.

This article  focusses on score based methods. A  recursive partitioning (tree based) method is proposed that allows to identify the items that carry DIF together with the variables that induce DIF. The variables can represent groups as in classical DIF detection techniques but can also include continuous variables like age. A strength of the method is that for continuous variables it is not necessary to define a priori the intervals that are relevant, the method itself generates the intervals that are linked to DIF. The resulting tree visualizes in a simple way the structure of DIF in an item showing which variables and interactions of variables generate DIF.

The method should be distinguished from the Rasch trees proposed by \citet{strobl2013rasch}. By using tree methodology the Rasch tree method also does not need pre-specified subgroups and can handle continuous variables. However, Rasch trees are IRT based methods. Moreover, they recursively partition the covariate space
to identify regions of the covariate space in which DIF occurs.  Regions are suspected to be relevant if the parameter estimates in the regions differ strongly. Therefore, regions in the covariate space are identified that show different difficulties but the method does not flag  items that are responsible. In contrast, the recursive partitioning method proposed here focusses on the detection of the items that are responsible for DIF. Recursive partitioning is used on the item level not on the global level, which treats all items simultaneously. The method is related to the recursive partitioning method proposed by \citet{TuBerg2015IF}. However, their method is IRT based and does not allow for non-uniform DIF.

In Section 2 we introduce the new recursive partitioning method based on the logistic regression approach for uniform DIF and in Section 3 we present an illustrative example. A detailed description of the fitting procedure is given in Section 4. In Section 5 we consider the results of various simulations. Models for the extension to non-uniform DIF are considered in Section 6. Finally, Section 7 contains two applications on real data.

\section{Logistic Regression Approaches to DIF} \label{sec:method}

In this section basic logistic regression approaches to the detection of uniform DIF are described and the alternative tree based method is introduced.

\subsection{Linear Logistic Regression Approaches to DIF}

The basic test score based method to detect uniform DIF was proposed by  \citet{swaminathan1990detecting}. It can be seen as a starting point of the method proposed here.

Let $Y_{pi} \in \{0,1\}$, $p=1,\hdots,P$, $i=1,\hdots,I$ denote the response when person $p$ tries to solve item $i$. \citet{swaminathan1990detecting} proposed to model the probability of solving an item as a function of the group membership and the test score by fitting the logistic regression model
\begin{equation}\label{eq:multiple_uniform}
\text{log}\left(\frac{P(Y_{pi}=1|S_p, g)}{P(Y_{pi}=0|S_p, g)}\right)=\eta_{pi}=\beta_{0i}+S_p\beta_i+\gamma_{ig},
\end{equation}
where $g$ denotes the group, $S_p$ is the test score of person $p$, $\beta_{0i}$ is the intercept, $\beta_i$ is the slope of item $i$ and $\gamma_{ig}$ are the  group-specific parameters.
In this model the parameters $\beta_{01}, \dots,\beta_{0I}$ represent the item difficulties and the parameters $\beta_{1}, \dots,\beta_{I}$ correspond to discrimination parameters. Within this framework the test scores are considered as proxies for the abilities of persons.
For the detection of DIF the most interesting parameters are the group-specific parameters $\gamma_{i1},\dots, \gamma_{iG}$, where $G$ denotes the number of groups. They represent the differential item functioning.
In the simplest case of two groups, a reference group and a focal group, one chooses $\gamma_{i1}=0$ for the reference group. Thus, for example, with groups defined by gender with female as the reference group one has
\begin{equation}\label{eq:multiple_gender}
\beta_{0i}+\gamma_{i,male}\;\text{ for males}\quad\text{and}\quad\beta_{0i}\;\text{ for females}.
\end{equation}
If $\gamma_{i,male} \ne 0$ one has DIF in item $i$ generated by gender. The original framework for two groups was proposed by \citet{swaminathan1990detecting}, the extension to multiple groups was considered by \citet{magis2011}. In the multiple group case one of the $G$ groups, for example the first group, has to be chosen as reference group by setting $\gamma_{i1}=0$.

DIF detection within the logistic regression framework typically uses likelihood ratio statistics that test the null hypothesis $H_0: \gamma_{i1}=\dots= \gamma_{iG}=0$. If the hypothesis is rejected item $i$ is considered as a DIF item. Each item is tested separately at significance level $\alpha$ with the degrees of freedom equal to $G-1$, depending on the number of groups.

The basic concept can be simply extended to  include   continuous (and categorical) variables that might induce DIF. Let $\xb_p^\top=(x_{p1},\hdots,x_{pm})$ be a vector of person-specific explanatory variables of length $m$.  An extension of model \eqref{eq:multiple_uniform} for uniform DIF has the form
\begin{equation}\label{eq:linear_uniform}
\text{log}\left(\frac{P(Y_{pi}=1|S_p,\xb_p)}{P(Y_{pi}=0|S_p,\xb_p)}\right)=\eta_{pi}=\beta_{0i}+S_p\beta_i+\xb_p^\top\gammab_i.
\end{equation}
The new intercept parameters in model \eqref{eq:linear_uniform} are $\beta_{0i}+\xb_p^\top\gammab_i$ and they differ according to the characteristics of the person $\xb_p$. The comparison of multiple groups is just a special case. Setting the first group as reference one defines the vector of explanatory variables $\xb_p^\top=(x_{p2},\hdots,x_{pG})$, where $x_{pg}=1$ if person $p$ is from group $g$ and 0 otherwise. The corresponding vector of parameters for one item $i$ is $\gammab_{i}^\top=(\gamma_{i2},\hdots,\gamma_{iG})$. Uniform DIF is present in this item if $\gammab_i\neq \0$. To investigate DIF one uses a global test for the whole parameter vector, $H_0:\gammab_i=0$. The alternative hypothesis is that at least one of the parameters are unequal to zero. The hypotheses are tested separately for each item at significance level $\alpha$. Due to the design of the tests the  approach identifies the items that carry DIF but does not contain any information about  the components of $\xb_p$ that are  responsible for DIF. Although being a straightforward extension of the fixed groups DIF model \eqref{eq:multiple_uniform} the extension \eqref{eq:linear_uniform} seems not to have been investigated so far.

We will refer to the multiple groups model \eqref{eq:multiple_uniform} as the \textit{classical} logistic regression modelling approach and to model \eqref{eq:linear_uniform} as the \textit{extended} approach.

\subsection{A Tree Representation of  DIF}

DIF detection based on the logistic regression model as described in the previous section has some limitations and drawbacks. If one uses the traditional version with $G$ groups DIF can be induced only by group membership. A continuous variable like age has to be divided into intervals to obtain groups without knowing which intervals are important. The extended version with a linear predictor is restricted by the assumption that the DIF effect is linear. Moreover, the tests that are used to identify items that carry DIF do not show which variables are responsible for DIF, at least not in a simple way. The proposed recursive partitioning method, for simplicity referred to as tree method, avoids the problem that reference and focal groups have to be specified a priori. By recursive splitting the method itself identifies the groups that induce DIF if they are present.

The general concept of recursive partitioning has its roots in automatic interaction detection. The most popular modern version is due to
\citet{BreiFrieOls:84} and is known by the name
\textit{classification and regression trees}, or
CART. An alternative approach is  the recursive partitioning framework based on conditional inference proposed by \citet{Hotetal:2006}. The basic method is
conceptually very simple. By binary recursive partitioning
 the feature space is partitioned
into a set of rectangles, and on each rectangle a simple model (for example, a constant) is fitted. An easily accessible  introduction into basic concepts is found in \citet{HasTibFri:2009B}, an overview with a focus on psychometrics was given by \citet{Strobetal:2009}. It should be noted that the method proposed here is based on the same idea but there is one crucial difference. When fitting a model we do not fit two separate models within the rectangles obtained by partitioning. We fit one closed model and only the intercept is partitioned into rectangles.
This yields item focussed trees in contrast to global trees as used by conventional Rasch trees.

Building a tree means to successively find a partition of the predictor space, where each node represents a subset of the predictor space. The terminal nodes of the tree build a disjoint partition of the predictor space and correspond to the relevant subregions of interest. When growing a tree one typically splits one node $A$ into two subsets $A_1$ and $A_2$. The split is determined by exactly one variable and the construction of the split depends on the scale of the variable. In the following considerations we will focus on metrically scaled and ordinal variables. In this case the partition  into two subsets has the form
\[
A_1 = A \cap \{ x_j \leq c \} \quad \text{and} \quad  A_2= A \cap \{ x_j > c \},
\]
with regard to threshold $c$ on variable $x_j$.
Given the covariates $\xb_p$ one can account for uniform DIF by building a partition of the respondents with differing intercepts. The first split with regard to the $j$-th variable and corresponding split point $c_j$ means to fit the model with predictor
\begin{equation}\label{eq:first_split}
\eta_{pi}=S_p\beta_i+[\gamma_{il}^{[1]}I(x_{pj} \le c_{j})+\gamma_{ir}^{[1]}I(x_{pj} > c_{j})],
\end{equation}
where $I(\cdot)$ denotes the indicator function with $I(a)=1$ if $a$ is true and $I(a)=0$ otherwise.  The parameter $\gamma_{il}^{[1]}$ denotes the intercept in the left node $(x_{pj} \le c_{j})$ and $\gamma_{ir}^{[1]}$ the intercept in the right node $(x_{pj} > c_{j})$. For example one split with regard to the binary covariate gender yields the intercepts
\[
\gamma_{il}^{[1]}=\gamma_{i,male}\;\text{ for males}\quad\text{and}\quad\gamma_{ir}^{[1]}=\gamma_{i,female}\;\text{ for females}.
\]
This parametrization is an equivalent representation of \eqref{eq:multiple_gender}. The main difference is that the two subgroups of interest are not predefined but determined by a split in variable $j$ at split-point $c_j$.
To determine the first split one examines all the null hypotheses $H_0:\gamma_{il}^{[1]}=\gamma_{ir}^{[1]}$. If $H_0$ cannot be rejected for any combination of variable and split point the item is considered to  be free of DIF. In the proposed algorithm likelihood ratio tests are used to examine the null hypotheses. In the very first step one chooses the combination of item, variable and split point with the smallest $p$-value of the corresponding test. If a significant effect is found the first split into left and right node is carried out for the selected item. In Section \ref{sec:algorithm} the splitting criterion is described in more detail.

One further split, for example in the right node $(x_{pj}>c_j)$, with regard to the $s$-th variable at split point $c_s$ yields the two daughter nodes $I(x_{pj} > c_{j})I(x_{ps} \le c_{s})$ and $I(x_{pj} > c_{j})I(x_{ps} > c_{s})$.
The new nodes are both defined by the product of two indicator functions. In general each node can be represented by a product of several indicator functions, namely
\[
node(\xb_p)=\prod_{b=1}^{B} I(x_{pj_b} > c_{j_b})^{a_{b}}I(x_{pj_b} \leq c_{j_b})^{1-a_{b}},
\]
where $B$ is the total number of indicator functions or branches, $c_{j_b}$ is the selected split point in variable $j_b$ and $a_b \in \{0,1\}$ indicates which of the indicator functions, below or above the threshold, is involved.
The resulting predictor of the model for item i after several splits with terminal nodes $\ell=1,\hdots,L_i$ is than given by
\begin{equation}\label{eq:whole_uniform}
\eta_{pi}=S_p\beta_i + \sum_{\ell=1}^{L_i}\gamma_{i\ell}\,node_{i\ell}(\xb_p)=S_p\beta_i +tr_i(\xb_p).
\end{equation}
where $tr_i(\xb_p)$ is the tree component containing subgroup-specific intercepts represented by the terminal nodes $node_{i\ell}(\xb_p)$.
The proposed algorithm yields an individual tree for each item that was selected to carry DIF. If an item is never chosen for splitting it is assumed to be free of DIF, and the fitted "tree" is a constant $tr_i(\xb_p)=\beta_{0i}$.

We use the abbreviation \textit{IFT} for item focussed trees based on the logistic regression framework.

\begin{figure}[!ht]
\centering
\includegraphics[width=1\textwidth]{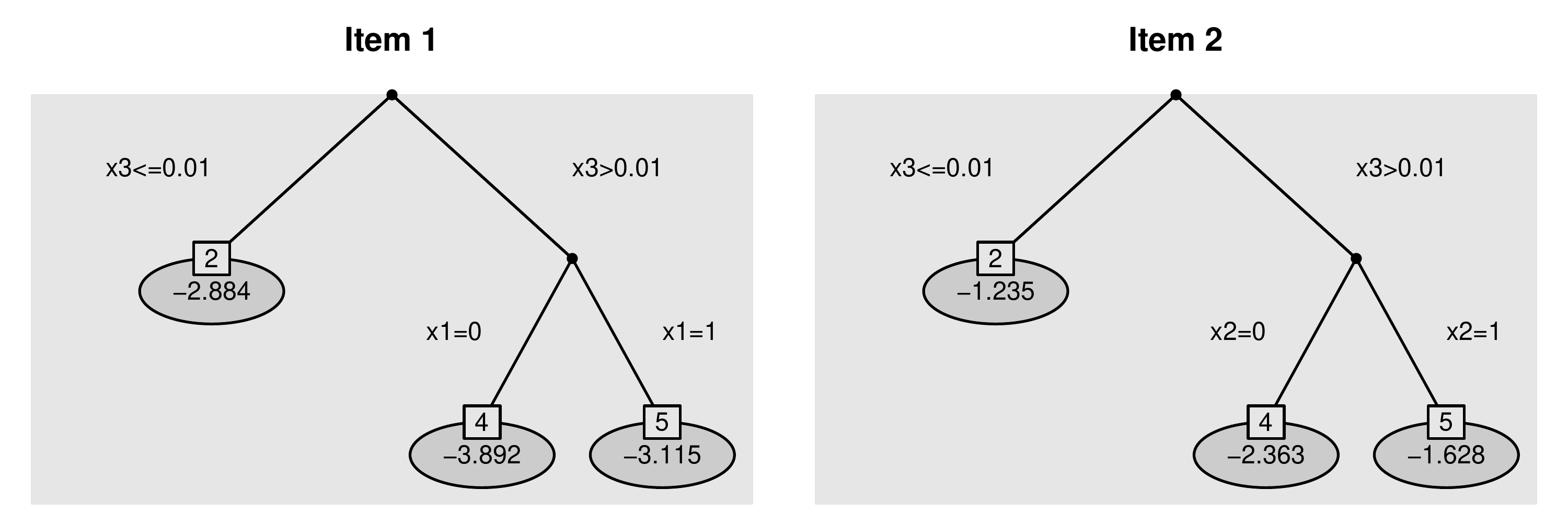}
\caption{Estimated trees of item 1 and 2 for the illustrative example. Estimated coefficients $\gamma_{i\ell}$ are given in each leaf of the trees.}
\label{fig:sim_uni3_example54}
\end{figure}

\section{An Illustrative Example}\label{sec:illustrative}

The procedure is now first illustrated by the use of artificial data. We consider data
$Y_{pi},\;p=1,\hdots,800,\;i=1,\hdots,20$, that are generated by a two-parameter model (2PL) with DIF. The basic 2PL model has the form
\begin{equation*}\label{eq:R}
P(Y_{pi}=1|\theta_p,b_i,a_i)=\frac{\text{exp}\left(a_i(\theta_p-b_i)\right)}{1+\text{exp}\left(a_i(\theta_p-b_i)\right)},
\end{equation*}
where $\theta_p$ denotes the person ability, $b_i$  the item difficulty and $a_i$ the item discrimination. We first generate person parameters $\theta_p$ and item difficulties $b_i$  from a standard normal distribution and item discriminations $a_i$ from a uniform distribution. However, instead of generating data from the 2PL model we assume that the difficulties of two of the $20$ items depend on covariates in a complex pattern.

In detail, we consider three covariates, two binary variables $x1,\,x2 \sim B(1,0.5)$ and one standard normal distributed variable $x3 \sim N(0,1)$.  In item 1 DIF is induced by $x1$ and $x3$ and the modified value of the difficulty is determined by the step functions $b_{1, \text{mod}}=b_1+ 0.8 \cdot I(x_3>0)+ 0.8 \cdot I\left(\{x_3>0\}\;\cap\;\{x_1=0\}\right)$, in item 2 DIF is induced by $x2$ and $x3$ and we use the step functions $b_{2, \text{mod}}=b_2+ 0.8 \cdot I(x_3>0) + 0.8 \cdot I\left(\{x_3>0\}\;\cap\;\{x_2=0\}\right)$, which represents an interaction between variables $x2$ and $x3$. In order to evaluate the fitting procedure 100 data sets were generated.

Figure \ref{fig:sim_uni3_example54} shows one exemplary estimation result of the two items with DIF (item 1 and 2) when fitting \textit{IFT}. The estimation in this example is quite perfect because the true underlying tree structure is detected for both items and no further item is falsely identified as DIF item. It can be seen from the trees that there are three groups represented by three terminal nodes, respectively. For item 1 it is distinguished between $\{x3\le 0.01\}$ and $\{x3>0.01\}$, and within this group between $\{x1=0\}$ and $\{x1=1\}$. The corresponding intercepts $\hat{\gamma}_{1\ell}$ and $\hat{\gamma}_{2\ell},\;\ell=1,\hdots,3$, of the estimated model \eqref{eq:whole_uniform} are given in each leaf of the trees.
According to model \eqref{eq:whole_uniform}, the probability to solve the item correctly increases with increasing intercepts. From the estimates in Figure \ref{fig:sim_uni3_example54} one can derive that item 1 is most difficult for region $\{x_3>0.01\}\;\cap\;\{x_1=0\}$ and item 2 is most difficult for $\{x_3>0.01\}\;\cap\;\{x_2=0\}$. These results are exactly in line with the true simulated effects. In the simulations in Section \ref{sec:sim} this artificial data is, inter alia, again considered in more detail.

\section{Fitting procedure} \label{sec:algorithm}

In this section we give details about the fitting procedure for our proposed tree-based model to investigate uniform DIF.

\subsection{Concepts}

When building trees for single items in each step one has to identify the best split due to an optimality criterion and decide if there is a relevance to perform the split or not. The second determines when to stop and therefore at the same time determines the size of the trees.

Since the approach is based on logistic regression models it is quite natural to use test based splits. In each step of the fitting procedure one obtains $p$-values for the two parameters that are involved in the splitting. In our previous notation one examines all the null hypotheses $H_0:\gamma_{il}=\gamma_{ir}$ for each combination of item, variable and split point. One simply selects the combination as the optimal one that has the smallest $p$-value. As test statistic we use the likelihood ratio (LR) test statistic. Computing the LR test statistic requires to estimate both models, the full model and the restricted model under $H_0$. We nevertheless prefer the LR statistic because it corresponds to select the model with minimal deviance. This criterion on the other hand is equivalent to minimizing the entropy, which belongs to the family of impurity measures that were already introduced as splitting criteria by \citet{BreiFrieOls:84}.

In order to decide if the split should be performed or not we use a concept based on maximally selected statistics. The idea is to perform a test that investigates the null hypotheses of independence of the response and one of the covariates at the global variable level. For one fixed item $i$ and variable $j$ one simultaneously considers all LR test statistics $T_{jc_j}$, where $c_j$ are from the set of possible split points, and computes the maximal value statistic $T_j=max_{c_j}T_{jc_j}$. The $p$-value that can be obtained by the distribution of $T_j$ provides a measure for the relevance of variable $j$. The result is not influenced by the number of split points, since it has already taken into account, see \citet{HotLau:03}, \citet{Shih:04}, \citet{ShiTsa:2004}, \citet{StrBouAug:2007}. As the distribution of $T_j$ in general is unknown we use a permutation test to obtain a decision on the null hypotheses. The distribution of $T_j$ is determined by computing the maximal value statistics based on random permutations of variable $j$. A random permutation of variable $j$ breaks the relation of the covariate and the response in the original data. By computing the maximal value statistics for a large number of permutations one obtains an approximation of the distribution under the null hypotheses and an corresponding $p$-value. All computations in the present article are based on 1000 permutations. Given overall significance level $\alpha$ the local significance level of one permutation test for fixed item and variable is chosen as $\alpha/m$. Using this adaption  the probability for each item without DIF of being falsely classified as DIF item is controlled by $\alpha$. As usual in DIF detection one controls for the type I error that is also known as false alarm rate. However, on the item level one should adapt for multiple testing. Choosing $\alpha/m$ ensures that the probability of falsely identifying at least one variable as responsible for DIF is controlled by $\alpha$.

\subsection{The Basic Algorithm}

The basic algorithm for uniform DIF is the following.

\vspace{0.5 cm}
\hrule
\begin{center}{\bf Basic Algorithm - Uniform DIF}\end{center}

\begin{description}
\item{\it Step 1 (Initialization)}

Set counter $\nu=1$

\begin{itemize}
\item[(a)] Estimation

For all items $i=1,\hdots,I$, fit all the candidate logistic models with predictor
\begin{align*}
\eta_{pi}= &S_p\beta_i+\gamma_{i1}I(x_{pj} \le c_{ijk})+\gamma_{i2}I(x_{pj} > c_{ijk}),\\
&j=1,\hdots,m,\quad k=1,\hdots,K_j
\end{align*}

\item[(b)] Selection

Select the model that has the best fit. Let $c_{i_1,j_1,k_1}$ denote the best split, which is found for item $i_1$ and variable $x_{j_1}$.

\item[(c)] Splitting decision

Select the item and variable with the largest value of $T_j$. Carry out permutation test for this combination with significance level $\alpha/m$. If significant, fit the selected model yielding estimates $\hat{\beta_i}$, $\hat{\gamma}_{i_1,1}$, $\hat{\gamma}_{i_1,2}$ and nodes $node_{i_1,1}, node_{i_1,2}$, set $\nu=2$. If not, stop, no DIF detected.
\end{itemize}

\item{\it Step 2 (Iteration)}

\begin{itemize}
\item[(a)] Estimation:

For all items $i=1,\hdots,I$ and already built nodes $\ell=1,\hdots,L_{i\nu}$, fit all the candidate logistic models with new intercepts
\[
\gamma_{i,L_{i\nu}+1}node_{i\ell}I(x_{pj}\leq c_{ijk})+\gamma_{i,L_{i\nu}+2}node_{i\ell}I(x_{pj}>c_{ijk})
\]
for all j and remaining, possible split points $c_{ijk}$.

\item[(b)] Selection

Select the model that has the best fit yielding the split point $c_{i_\nu,j_\nu,k_\nu}$, which is found for item $i_\nu$ in node $node_{i_\nu,\ell_\nu}$ and variable $x_{j_\nu}$

\item[(c)] Splitting decision

Select the node and variable with the largest value of $T_j$. Carry out permutation test for this combination with significance level $\alpha/m$. If significant, fit the selected model yielding the additional estimates $\hat{\gamma}_{i_\nu,L_{i_\nu,\nu}+1}, \hat{\gamma}_{i_\nu,L_{i_\nu,\nu}+2}$, set $\nu=\nu+1$. If not, stop.

\end{itemize}
\end{description}
\hrule
\vspace{0.5 cm}

\section{Simulations} \label{sec:sim}

In the following we consider data $Y_{pi},\;p=1,\hdots,P,\;i=1,\hdots,I$ that are generated according to the two-parameter model (2PL), which is a dichotomous IRT model of the form

\begin{equation}\label{eq:2PL}
P(Y_{pi}=1|\theta_p,a_i,b_i)=\frac{\text{exp}\left(a_i(\theta_p-b_i)\right)}{1+\text{exp}\left(a_i(\theta_p-b_i)\right)},
\end{equation}
where $\theta_p$ are the person abilities, $b_i$ are the item difficulties and $a_i$ are the item discrimination parameters.

We consider several simulation scenarios where in a first step the person parameters $\theta_p$ and the item difficulties $b_i$ are independently drawn from a standard normal distribution and the item discrimination parameters $a_i$ are  uniformly distributed, $a_i \sim U(0,1)$. If an item $i$ is assumed to show uniform DIF the corresponding parameter $b_i$  is subsequently transformed by specific step functions in each scenario. A detailed description is given in the respective section.

In each simulation scenario we vary the number of persons, $P\in\{400,800\}$, the number of items, $I\in\{20,40\}$, and the percentage of DIF items, which is $0\%$, $10\%$ or $20\%$. In the cases with DIF we additionally consider two different strengths of DIF, given for each scenario in the respective section. In total this results in 20 different settings (4 without DIF and 16 with DIF) respectively. In each setting 100 data sets were generated. During estimation each permutation test is based on 1000 permutations.

In order to evaluate the performance of the proposed tree based model \eqref{eq:whole_uniform}
we compute true positive rates (TPR), also named hit rates, and false positive rates (FPR), which correspond to the Type I error rates if no DIF is present. We distinguish between TPR and FPR on the item level and for the combination of item and variable.
Let each item be characterized by a vector $\deltab_i^T=(\delta_{i1},\hdots,\delta_{im})$, where $m$ denotes the number of covariates, with $\delta_{ij}=1$ if item $i$ has DIF in variable $j$ and $\delta_{ij}=0$ otherwise. An item is a non-DIF item if $\deltab_i^T=(0,\hdots,0)$, if one of the components is 1 it is a DIF item. With indicator function $I(\cdot)$, the criteria to judge the identification of items with DIF are:

\begin{itemize}
\item[-] True positive rate on the item level:

$TPR_I=\frac{1}{\#\{i:\deltab_i\neq \mathbf{0}\}}\sum_{i:\deltab_i\neq \mathbf{0}}{I(\hat{\deltab}_i\neq\mathbf{0})}$

\item[-] False positive rate on the item level:

$FPR_I=\frac{1}{\#\{i:\deltab_i=\mathbf{0}\}}\sum_{i:\deltab_i= \mathbf{0}}{I(\hat{\deltab}_i\neq\mathbf{0})}$

\item[-] True positive rate for the combination of item and variable:

$TPR_{IV}=\frac{1}{\#\{i,j:\delta_{ij}\neq 0\}}\sum_{i,j:\delta_{ij}\neq 0}{I(\hat{\delta}_{ij}\neq 0)}$

\item[-] False positive rate for the combination of item and variable:

$FPR_{IV}=\frac{1}{\#\{i,j:\delta_{ij}= 0\}}\sum_{i,j:\delta_{ij}= 0}{I(\hat{\delta}_{ij}\neq 0)}$.
\end{itemize}
The methods that are considered in the simulations are
\begin{itemize}
\item[-] \textit{Logistic}, which denotes the \textit{classical} regression method proposed by \citet{swaminathan1990detecting} and \citet{magis2011}. If the predictor is a vector with possibly continuous variables it denotes the \textit{extended} logistic model.
\item[-] \textit{IFT} for item focussed trees based on the logistic model, which describes the recursive partitioning method proposed here.
\end{itemize}

\subsection{Results}

First
we consider data with two or more groups defined by one covariate. The main objective here is to compare the proposed \textit{IFT} approach to the classical \textit{Logistic} approach, which is well established for the comparison of multiple groups. Later we give detailed results of the proposed \textit{IFT} considering more complex data constellations with several predictors.

\subsubsection{One binary predictor} \label{subsubseq:sim_uni_binaer}

\begin{figure}[!ht]
\centering
\includegraphics[width=0.75\textwidth]{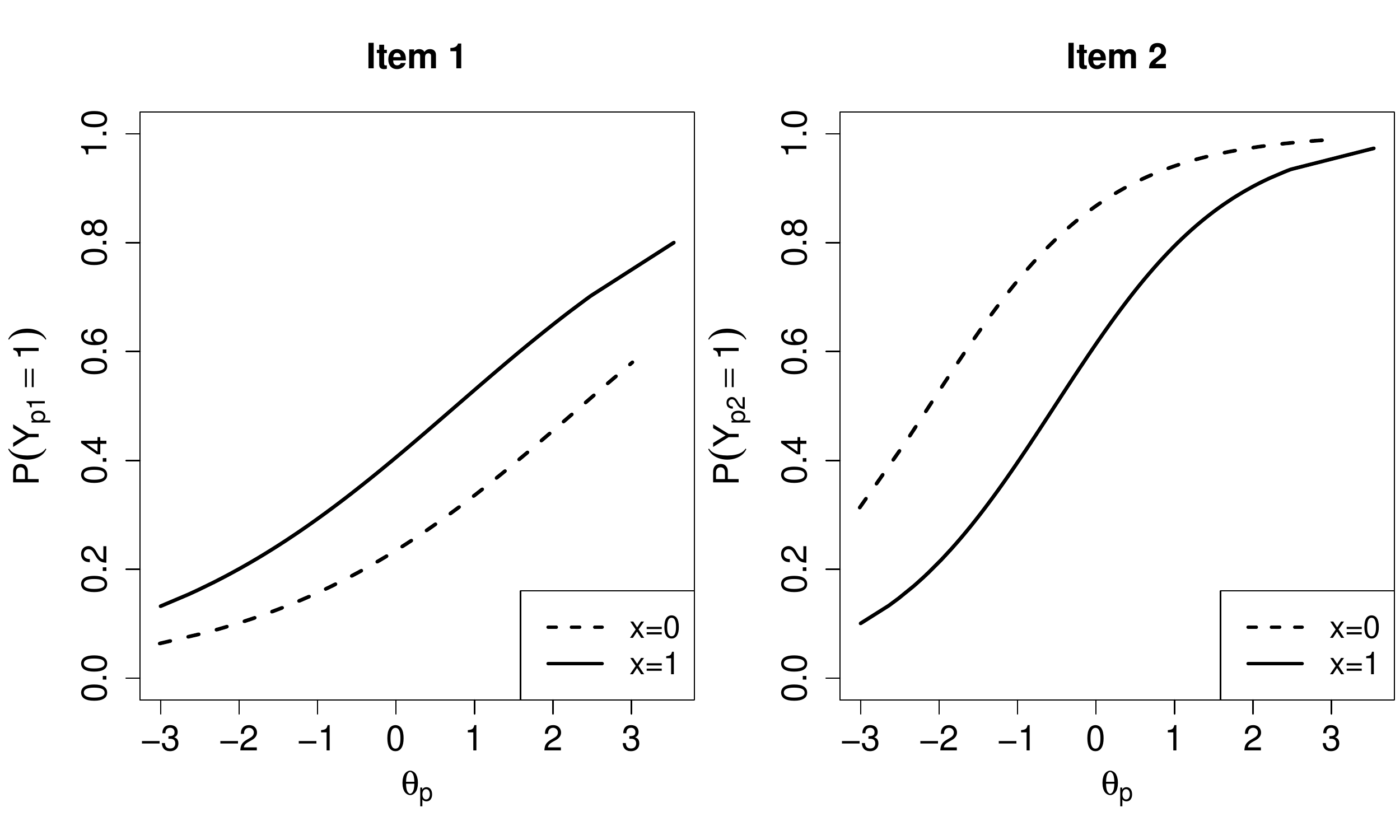}
\caption{Item Characteristic Curves of item 1 and item 2 for one setting in the simulation with one binary predictor.}
\label{fig:sim_uni1_ICC}
\end{figure}

We start with one binary covariate $x\in\{0,1\}$. In this simple case the investigations reduce to the comparison of two groups. Uniform DIF is present if the item difficulties $b_i$ differ between the two groups. The difference is simulated by $b_{i,\text{mod}} = b_i + c \cdot I(x=0)$ for one half of the DIF items and $b_{i,\text{mod}} = b_i + c \cdot I(x=1)$ for the other half of the DIF items. The strength of DIF is determined by the constant $c\in\{0.8,1.6\}$. DIF is generated symmetrically because one half of DIF items favour the first group ($x=1$) and the other DIF items favour the second group ($x=0$).
For illustration Figure \ref{fig:sim_uni1_ICC} shows the Item Characteristic Curves (ICC) of the two items with DIF for the setting with $P=800$, $I=20$, $10\%$ DIF items and $c=1.6$. From the probabilities it can be seen that item 1 is more difficult for $x=0$ and item 2 is more difficult for $x=1$. The item locations (value of $\theta_p$ with probability 0.5) differ between the two groups but the item discriminations (steepness at the item location) are the same for both groups.

\begin{figure}[!ht]
\centering
\includegraphics[width=0.9\textwidth]{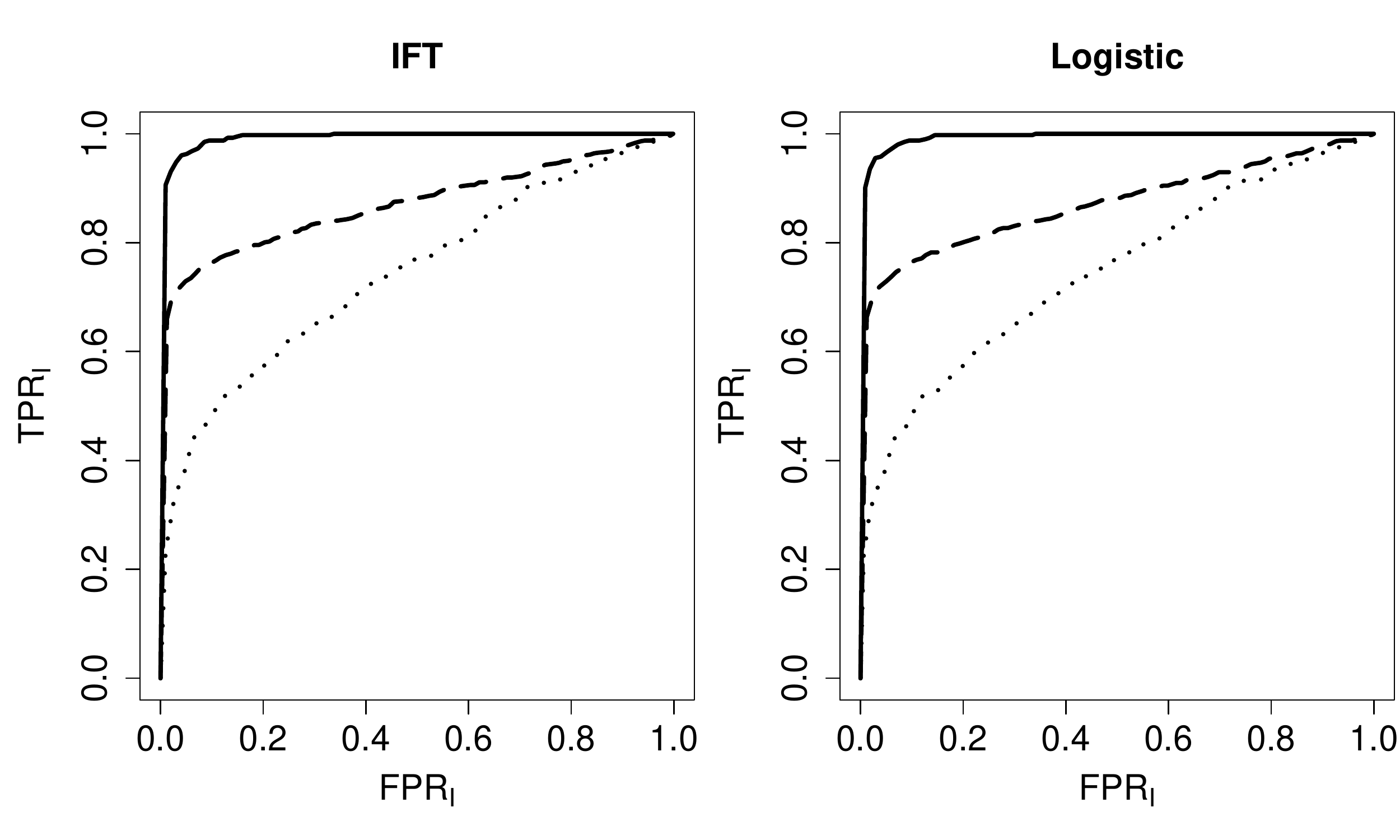}
\caption{Average ROC curves for three settings (different linetypes) in the simulation with one binary predictor.}
\label{fig:sim_uni1_ROC_3}
\end{figure}

For the comparison of the results we use Receiver Operating Characteristic (ROC) curves, which have also been used by \citet{magis2014detection} and \citet{SchauTuBoost2015} to evaluate the performance of DIF detection methods. True positive rates and false positive rates on the item level were computed for increasing significance level $\alpha \in \rbrack 0,1 \lbrack $. The corresponding ROC curve is then obtained by plotting $(FPR_I,TPR_I)$ as a function of $\alpha$. Figure \ref{fig:sim_uni1_ROC_3} shows the ROC curves for three out of 16 settings with DIF as the average over 100 repetitions, respectively. The left panel shows the resulting curves for \textit{IFT} and the right panel shows the resulting curves for the classical \textit{Logistic} method. The solid line in Figure \ref{fig:sim_uni1_ROC_3} corresponds to the setting with $P=400$, $I=40$, $10\%$ DIF items and $c=1.6$, the dashed line corresponds to the setting with $P=800$, $I=40$, $20\%$ DIF items and $c=0.8$ and the dotted line corresponds to the setting with $P=400$, $I=20$, $20\%$ DIF items and $c=0.8$.

\begin{table}[!ht]
\caption{Average FPR on the item level at significance level $\alpha=0.05$ for the four settings without DIF in the simulation with one binary predictor.}
\begin{center}
\begin{tabularsmall}{lcccc}
\toprule
FPR$_I$&\multicolumn{2}{c}{I=20}&\multicolumn{2}{c}{I=40}\\
&P=400&P=800&P=400&P=800\\
\midrule
IFT&0.050&0.051&0.049&0.050\\
Logistic&0.052&0.048&0.051&0.050\\
\bottomrule
\end{tabularsmall}
\end{center}
\label{tab:sim_uni1_FPR_0}
\end{table}

Although the global performance  varies over the different settings, there are only minor differences between  the two methods as far as their performance is concerned. All settings we considered, not only the one presented in  Figure \ref{fig:sim_uni1_ROC_3}, showed nearly no differences between the two methods. This result is not really surprising. After one split in to the binary predictor $x$ the obtained model \eqref{eq:whole_uniform} for one item is exactly the same as model \eqref{eq:linear_uniform}, which is used for testing when using the classical \textit{Logistic} approach. In this case the only remaining difference  is the use of different test statistics to obtain a decision. Nevertheless, the classical and the new approach obviously show the same performance. This is important because  the tree based approach, which can also be used in more complex settings with many variables, can also be used in the case of two groups without loss of efficiency.

The construction of ROC curves is an efficient tool but is informative only if  DIF is present. Therefore, we separately consider the case without DIF. The average false positive rates with significance level $\alpha=0.05$ for the four settings without DIF are given in Table \ref{tab:sim_uni1_FPR_0}. The absence of DIF is a baseline situation to check a possible inflation of false positive rates. According to the obtained results this is not the case. The \textit{IFT} approach (approximately) holds the significance level as does the classical \textit{Logistic} approach. Again, the two approaches nearly yield the same results.

\begin{figure}[!ht]
\centering
\includegraphics[width=0.65\textwidth]{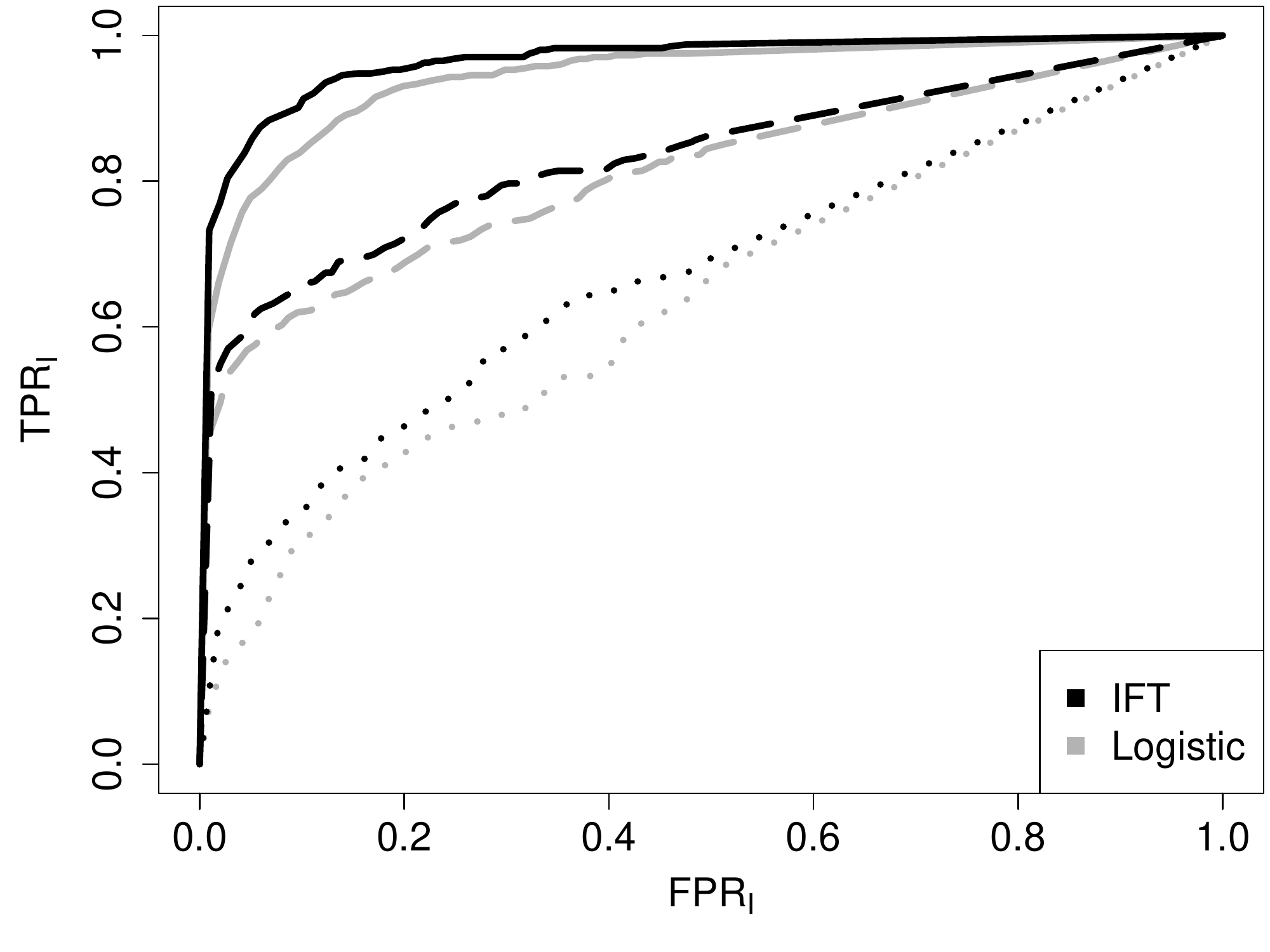}
\caption{Average ROC curves for three settings (different linetypes) in the simulation with one ordered predictor.}
\label{fig:sim_uni1b_ROC_3}
\end{figure}

\subsubsection{One Ordered predictor} \label{subsubseq:sim_uni_ordered}

Here we consider an ordered factor $x \in \{1,\hdots,6\}$. The difference in item difficulties is simulated by $b_{i, \text{mod}} = b_i + c \cdot I(x>3)$ for one half of DIF items and $b_{i, \text{mod}} = b_i + c \cdot I(x\le 3)$ for the other half of DIF items. Hence there are only two groups that show a true difference, respectively. All the other specifications remain the same as in the previous section \ref{subsubseq:sim_uni_binaer}. The ROC curves of three selected examples are given in Figure \ref{fig:sim_uni1b_ROC_3}. The chosen settings are different from those in Figure \ref{fig:sim_uni1_ROC_3}. The solid lines belong to the setting with $P=800$, $I=20$, $20\%$ DIF, $c=0.8$, the dashed lines to the setting with $P=400$, $I=20$, $20\%$ DIF, $c=1.6$ and the dotted lines belong to the setting with $P=400$, $I=20$, $10\%$ DIF, $c=0.8$.

In contrast to the comparison of two groups, now there are visible differences between the performances of the two methods. The ROC curves show that \textit{IFT} (black lines) outperforms the classical \textit{Logistic} (grey lines) across the whole range of $\alpha$. The ROC curves of the new approach always are everywhere above the ROC curves of the classical approach. These results are consistent throughout all settings. The differences are even stronger for the settings with weaker DIF $(c=0.8)$.
The reason for the better performance of \textit{IFT} is that it is able to use the ordering of the categories. Since DIF is linked to the ordinal scale of the factor a method that is able to exploit the ordering should perform better than the classical method that just distinguished between the groups.

\subsubsection{Several Predictors}

In the following simulations we consider three covariates, two binary variables $x1,\,x2 \sim B(1,0.5)$ and one standard normal distributed variable $x3 \sim N(0,1)$. Since \textit{IFT} allows to determine the variables that are responsible for DIF, true positive and false positive rates for the combination of item and variable can be computed. In the following all the presented results are based on computations with significance level $\alpha=0.05$. To account for the three covariates in the model the local significance level for one permutation test is $0.05/3$.

Before simulating items with DIF we first investigate the baseline situation without DIF. The average false positive rates for the four settings (varying number of persons and items) without DIF are given in Table \ref{tab:sim_uni2_FPR_0}. It is seen that \textit{IFT} yields small false positive rates. The procedure  is somewhat conservative and does not fully use the specified significance level. On average only one item is misleadingly identified as DIF item. False positive rates for the combination of item and variable are much smaller. With 40 items the value $0.008$ means that only one split with regard to a variable that was not inducing DIF was falsely executed during estimation.

\begin{table}[!ht]
\caption{Average FPR at significance level $\alpha=0.05$ for the four settings without DIF in the simulation with three covariates.}
\begin{center}
\begin{tabularsmall}{llcccc}
\toprule
&&\multicolumn{2}{c}{I=20}&\multicolumn{2}{c}{I=40}\\
&&P=400&P=800&P=400&P=800\\
\midrule
\multirow{2}{*}{IFT}&$FPR_I$&0.027&0.021&0.024&0.022\\
&$FPR_{IV}$&0.009&0.007&0.008&0.007\\
\bottomrule
\end{tabularsmall}
\end{center}
\label{tab:sim_uni2_FPR_0}
\end{table}

\begin{figure}[!ht]
\centering
\includegraphics[width=1\textwidth]{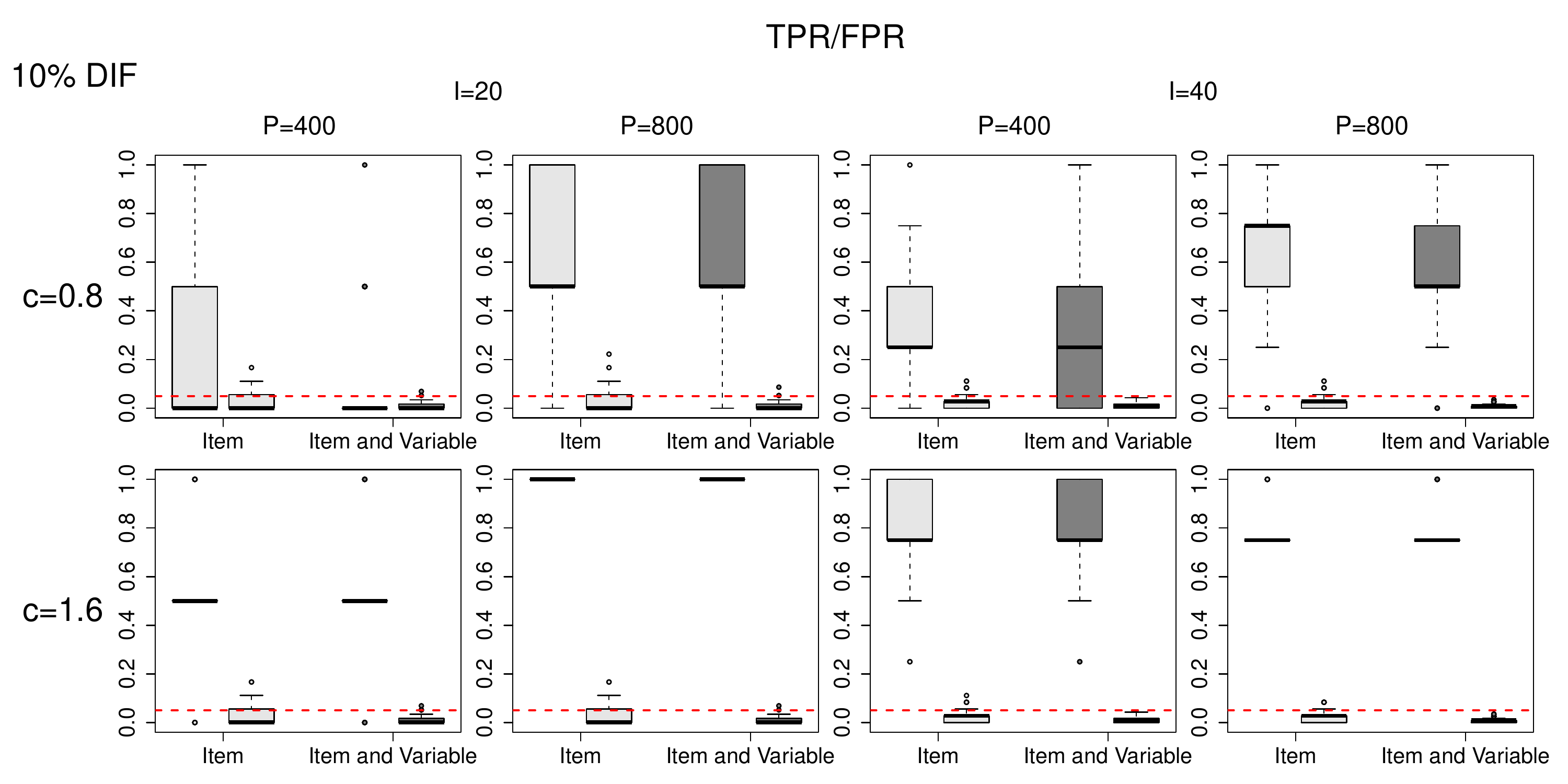}

\vspace{0.5cm}

\includegraphics[width=1\textwidth]{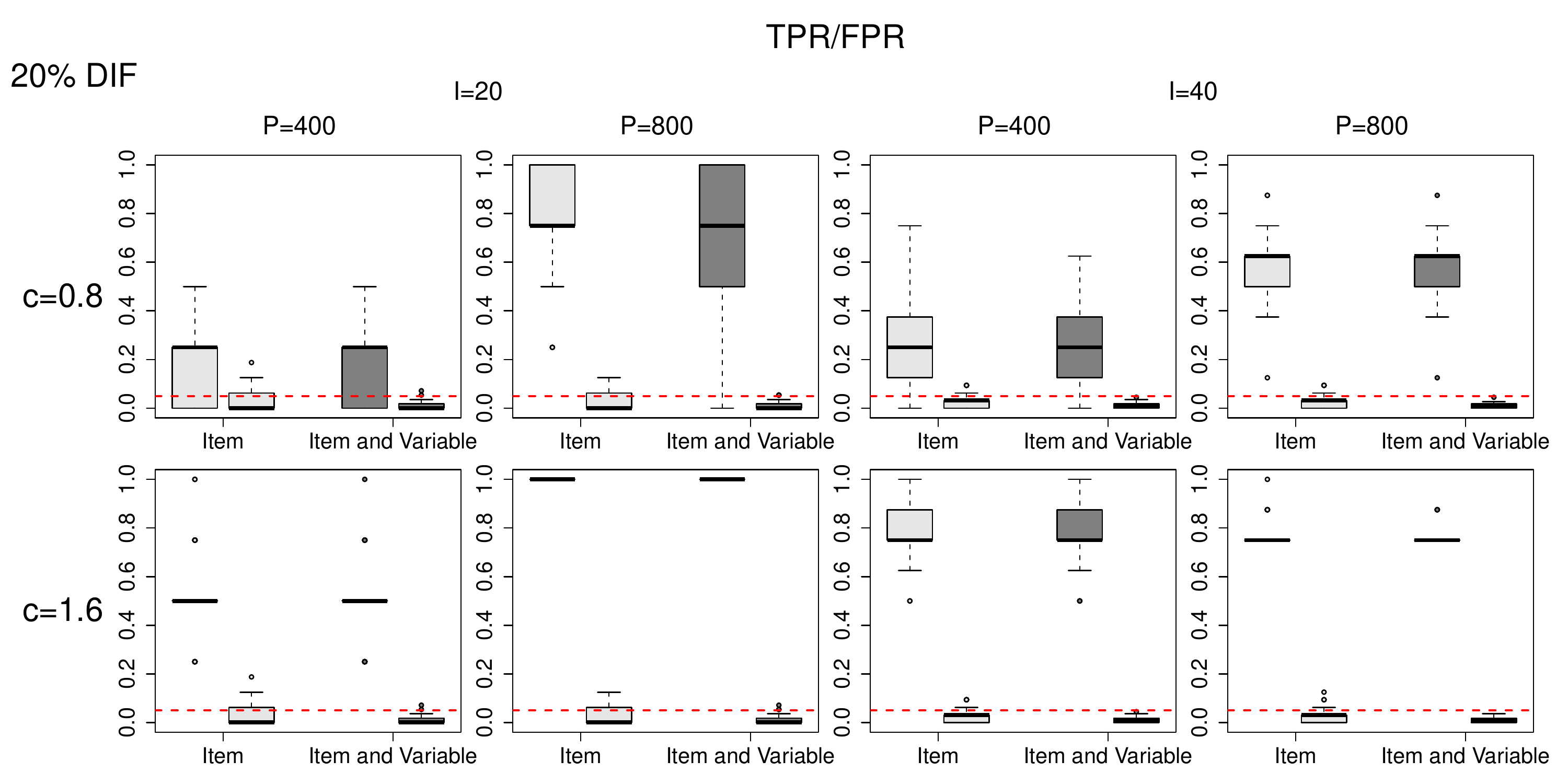}
\caption{Boxplots of TPR and FPR at significance level $\alpha=0.05$ (marked by dashed lines) in the simulation with three covariates and DIF in $x1$. Results on item level are given in light grey, results for the combination of item and variable are given in dark grey.}
\label{fig:sim_uni2_TPFP12}
\end{figure}

\subsubsection*{DIF in the First Variable}

In the settings with  DIF,  first DIF is simulated as in the simulation with one binary predictor only (Section \ref{subsubseq:sim_uni_binaer}). If DIF is present, the item difficulties $b_i$ differ between the two groups defined by the binary covariate $x1$. Hence the underlying true model is  defined by one split in $x1$.
Boxplots of true positive and false positive rates of the 16 settings with DIF are given in Figure \ref{fig:sim_uni2_TPFP12}. The results on the item level are  in light grey and are given on the left of each panel, the results for the combination of item and variable are  in dark grey and are given on the right of each panel. In addition, the significance level $\alpha=0.05$ is marked as a reference by dashed lines. It is seen from Figure \ref{fig:sim_uni2_TPFP12} that \textit{IFT} shows  good overall performance, in particular if the number of persons is large. True positive rates are high, especially in the settings with $P=800$ and $c=1.6$. False positive rates are very small throughout all settings, in particular the global significance level holds (with a tendency of the method to be conservative). It is  noteworthy that the true positive rates for the combination of item and variable in all settings are very similar to the true positive rates for items. Therefore,  \textit{IFT} is  able to simultaneously identify the items and variables that are responsible for DIF. It should be noted that in classical approaches for fixed groups the simultaneous detection of DIF item and responsible variable is not investigated. If one  considers more than one categorical variable, for example, gender and race, typically DIF induced by gender and race are investigated separately with significance levels fixed to the same value separately for the two investigations.

\begin{figure}[!ht]
\centering
\includegraphics[width=1\textwidth]{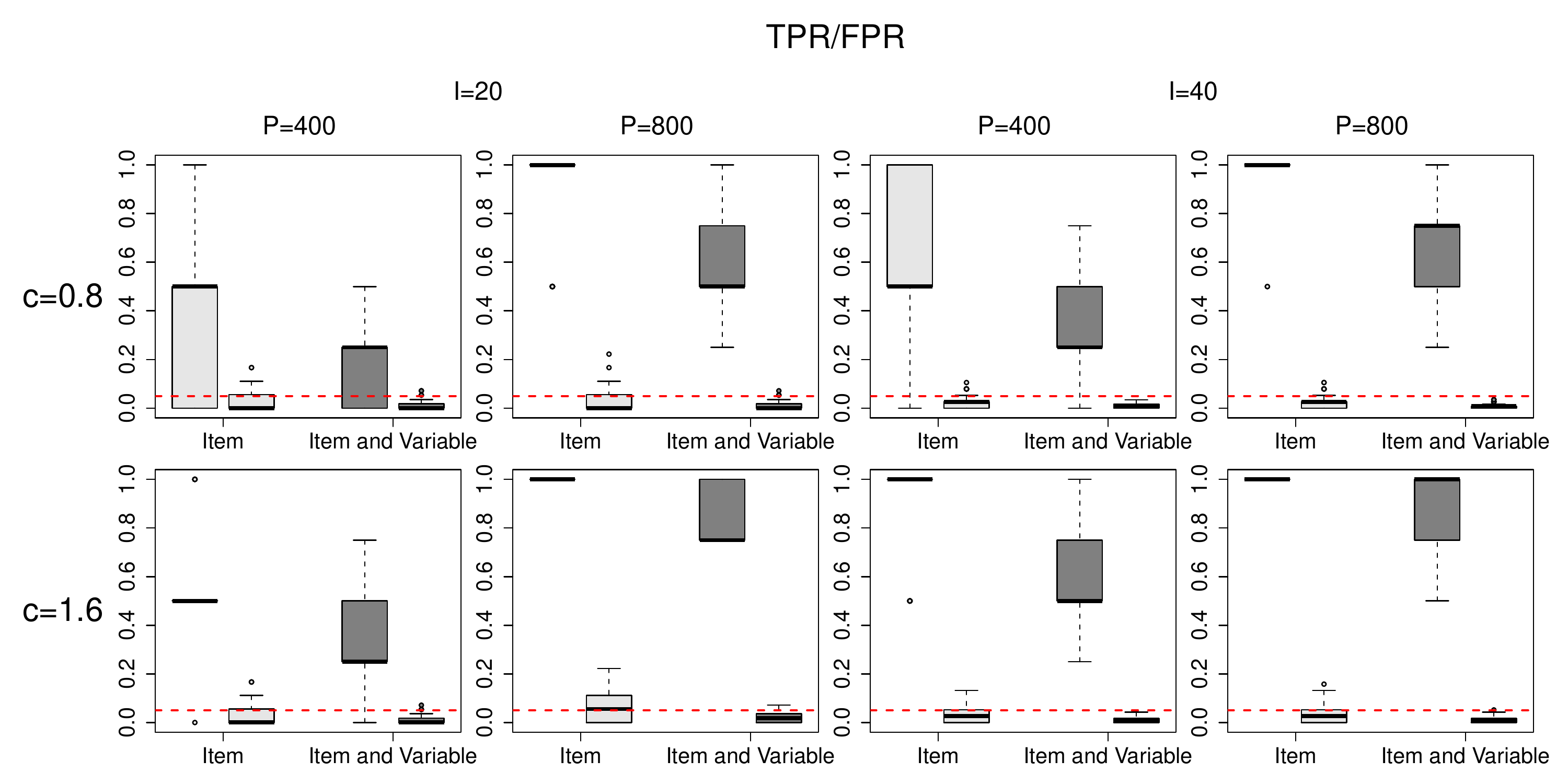}
\caption{Boxplots of TPR and FPR at significance level $\alpha=0.05$ (marked by dashed lines) in the simulation with three covariates and DIF in two items and two covariates. Results on item level are given in light grey, results for the combination of item and variable are given in dark grey.}
\label{fig:sim_uni3_TPFP}
\end{figure}

\subsubsection*{DIF in two covariates}

In the following we consider again the complex DIF structure considered in the illustrative example and use two DIF items. In item 1 DIF is induced by $x1$ and $x3$ and determined by the step functions $b_{1, \text{mod}}=b_1+c\cdot I(x_3>0)+c\cdot I\left(\{x_3>0\}\;\cap\;\{x_1=0\}\right)$, in item 2 DIF is induced by $x2$ and $x3$ and we use the step functions $b_{2, \text{mod}}=b_2+ c \cdot I(x_3>0) + c \cdot I\left(\{x_3>0\}\;\cap\;\{x_2=0\}\right)$. The strength of DIF again is determined by the additional parameter $c\in\{0.8,1.6\}$. The choice of this values for c effects that the differences between the individual groups remain the same as in the previous simulations.

In the same way as in Figure \ref{fig:sim_uni2_TPFP12}, the true positive rates and false positive rates of the eight settings (with varying $I$, $P$ and $c$) based on 100 replications are given in Figure \ref{fig:sim_uni3_TPFP}. The true positive rates on the item level (given in light grey) are very high for all settings. Especially for the settings with $P=800$ the selection of items is quite perfect. It is also seen that the hit rates for the combination of item and variable (given in dark grey) are not so much smaller than the hit rates for items. Since here DIF is generated by two variables \textit{IFT} cannot detect both variables in all the cases. However, the small false positive rates show that the procedure does not tend to perform splits with regard to variables that are not responsible for DIF. If a significant effect is found the corresponding split is always in the right variable.

\section{Investigation of Non-Uniform DIF}

A strength of the logistic framework for DIF detection proposed by \citet{swaminathan1990detecting} is that it can be extended to detect non-uniform DIF.
We first consider the classical and extended approach and then the tree based method.

\subsection{Logistic Regression  for Non-Uniform DIF}
Let us again first consider the comparison of multiple groups. To account for non-uniform DIF model \eqref{eq:multiple_uniform} has to be extended by group-specific slopes and has the form
\begin{equation}\label{eq:multiple_nonuniform}
\eta_{pi}=\beta_{0i}+S_p\beta_i+\gamma_{ig}+S_p\alpha_{ig}
\end{equation}
where $\alpha_{ig}$ are the additional group-specific slopes. The first group is chosen as reference group by setting $\gamma_{i1}=\alpha_{i1}=0$, see, for example, \citet{magis2011}.
The model can be extended to account for non-uniform DIF that is  generated by a vector of covariates in a similar way as for uniform DIF.
Then one uses the model
\begin{equation}\label{eq:linear_nonuniform}
\eta_{pi}=\beta_{0i}+S_p\beta_i+\xb_p^\top\gammab_i+S_p\xb_p^\top\alphab_i,
\end{equation}
which contains an interaction between the person characteristics $\xb_p$ and the test score $S_p$.
The new slope parameters in model \eqref{eq:linear_nonuniform} are contained in $S_p(\beta_i+\xb_p^\top\alphab_i)$.
Model \eqref{eq:linear_nonuniform} reduces to the logistic model used in Section 2 if $\alphab_i=\0$. Thus uniform DIF is present if $\gammab_i\neq\0$ given $\alphab_i=\0$.
However, the item shows non-uniform DIF  if $\alphab_i\neq \0$ whether $\gammab_i=\0$ or not.


\subsection{Logistic Regression Trees for Non-Uniform DIF}
When using the proposed tree based model, non-uniform DIF means that splits are not only admissible in the variables $x_{p1},\hdots,x_{pm}$, but also in the interaction terms $S_px_{p1},\hdots,S_px_{pm}$. A (first) split with regard to the interaction between the test score and the $j$-th variable yields the model with predictor
\begin{equation*}\label{equ:lognonu}
\eta_{pi}=\beta_{0i}+S_p\;[\alpha_{il}^{[1]} I(x_{pj} \le c_{j})+\alpha_{ir}^{[1]} I(x_{pj} > c_{j})],
\end{equation*}
where the parameter $\alpha_{il}^{[1]}$ denotes the slope in the left node $(x_{pj} \le c_{j})$ and  $\alpha_{ir}^{[1]}$ denotes the slope in the right node $(x_{pj} > c_{j})$.

\subsection{Test Strategies}

In the literature different strategies were proposed how to test for the significance of DIF by means of model \eqref{eq:multiple_nonuniform}, see, for example, \citet{zumbo1999handbook} and \citet{magis2011}. We will  use similar strategies when testing for DIF in the extended logistic regression model \eqref{eq:linear_nonuniform} and the tree based approach.

\subsubsection*{Testing for DIF}
The first strategy is to test for both types of DIF effects simultaneously. The corresponding null hypothesis given model \eqref{eq:multiple_nonuniform} is $H_0:\gamma_{i2}=\hdots=\gamma_{iG}=\alpha_{i2}=\hdots=\alpha_{iG}=0$. For model \eqref{eq:linear_nonuniform} the corresponding null hypothesis is given by $H_0:\gammab_i=\alphab_i=0$. That means  DIF is investigated by using  a global test for the whole parameter vector $(\gammab_i,\alphab_i)$. DIF is considered as being present (in any form) if the test rejects the null hypothesis, meaning that at least one of the parameters $\gamma_{ij},\,\alpha_{ij},\;j=1,\hdots,m$,  differs from zero.

For item focussed trees the equivalent is  that at least one split is performed in one of the components. When selecting the optimal split in each step of the algorithm, one has to consider all combinations of item, variable, split point and component with regard to intercept and slope. The final model consists of one or two separate trees, one referring to the intercept and one referring to the slope. In general the trees will be different but can also have the same structure. The resulting tree
is given by
\begin{equation}\label{eq:whole_nonuniform}
\eta_{pi}=tr_i(\xb_p)+tr_i(S_p,\xb_p),
\end{equation}
where $tr_i(\xb_p)$ is the tree component containing subgroup-specific intercepts and $tr_i(S_p,\xb_p)$ is the tree component containing subgroup-specific slopes. In contrast to the tree in model \eqref{eq:whole_uniform} for uniform DIF now one has two possible trees. If there is only a significant effect in one of the two components a constant $tr_i(\xb_p)=\beta_{0i}$ or $tr_i(S_p,\xb_p)=S_p\beta_i$ is fitted in the other component.

In comparison to the classical and extended \textit{Logistic} method, the tree based model has two advantages:
\begin{itemize}
\item[-] The  obtained tree(s) distinguish between items with uniform and non-uniform DIF. The trees themselves show which form of DIF is present. Thus both types of DIF can be detected simultaneously within one fitting procedure.
\item[-] The obtained tree(s) identify the variables that induce uniform and/or non-uniform DIF. In particular, both types of DIF can be caused by different variables.
\end{itemize}

\subsubsection*{Testing for Non-Uniform DIF}
A second strategy is to explicit test for non-uniform DIF. Using the extended Logistic model \eqref{eq:linear_nonuniform} one investigates the null hypothesis $H_0:\alphab_i=\0$ for each item.
Non-uniform DIF is considered as being present if the hypothesis is rejected, meaning that at least one parameter $\alpha_{ij}$  differs from zero.

For item focussed trees the detection of non-uniform DIF  means that a significant split in the \textit{slope} component is found. Consequently, during estimation only the models with \textit{simultaneous splits} in the intercepts and the slopes are considered as potential candidates.
Therefore, one split in item $i$ with regard to variable $j$ corresponds to the model with predictor
\begin{equation}\label{eq:first_split_non}
\eta_{pi}=[\gamma_{il}^{[1]}I(x_{pj} \le c_{j})+\gamma_{ir}^{[1]}I(x_{pj} > c_{j})]+S_p\;[\alpha_{il}^{[1]} I(x_{pj} \le c_{j})+\alpha_{ir}^{[1]} I(x_{pj} > c_{j})],
\end{equation}
which contains  two intercepts $(\gamma_{il}^{[1]},\gamma_{ir}^{[1]})$ and  two slopes $(\alpha_{il}^{[1]},\alpha_{ir}^{[1]})$ with respect to the same subgroups. To select the optimal split and to determine the splitting decision one compares the likelihoods of model \eqref{eq:first_split} and \eqref{eq:first_split_non}. The procedure is continued in each step of the algorithm, considering all combinations of item, variable and split point.

If non-uniform DIF is present, the final model consists of two trees containing subgroup-specific intercepts and subgroup specific slopes  that are determined by the same splits.
\\[1em]
\noindent For the different strategies we will use the same terminology as \citet{magis2011} in his investigation of the case in which DIF is induced by multiple groups:
\begin{itemize}
\item[-] \textit{UDIF}  means testing for uniform DIF, $H_0:\gammab_i=0$, given model \eqref{eq:linear_uniform} within the logistic regression approach. For trees it refers to testing the corresponding splits.
\item[-] \textit{DIF} means simultaneous tests for uniform and non-uniform DIF, $H_0:\gammab_i=\alphab_i=0$, given model \eqref{eq:linear_nonuniform} for logistic regression. For trees it refers to testing the corresponding splits for both types of DIF.
\item[-] \textit{NUDIF} means  tests for  non-uniform DIF, $H_0:\alphab_i=\0$, given model \eqref{eq:linear_nonuniform} for logistic regression. For trees it refers to testing the corresponding splits.
\end{itemize}

\begin{table}[!ht]
\caption{Modified values of item discrimination and item difficulty parameters in the illustrative example with non-uniform DIF.}
\begin{center}
\begin{tabularsmall}{llcll}
\toprule
Item&Non-Uniform DIF&&Item&Uniform DIF\\
\midrule
1&$a_{1, \text{mod}}=a_1+ 0.6 \cdot I(x_1=1)$&&3&$b_{3, \text{mod}}=b_3+ 0.8 \cdot I(x_1=1)$\\
2&$a_{2, \text{mod}}=a_2+ 0.6 \cdot I(x_2=0)$&&4&$b_{4, \text{mod}}=b_4+ 0.8 \cdot I(x_2=0)$\\
\bottomrule
\end{tabularsmall}
\end{center}
\label{tab:example_nonuni_sf}
\end{table}

\begin{figure}[!ht]
\centering
\includegraphics[width=0.75\textwidth]{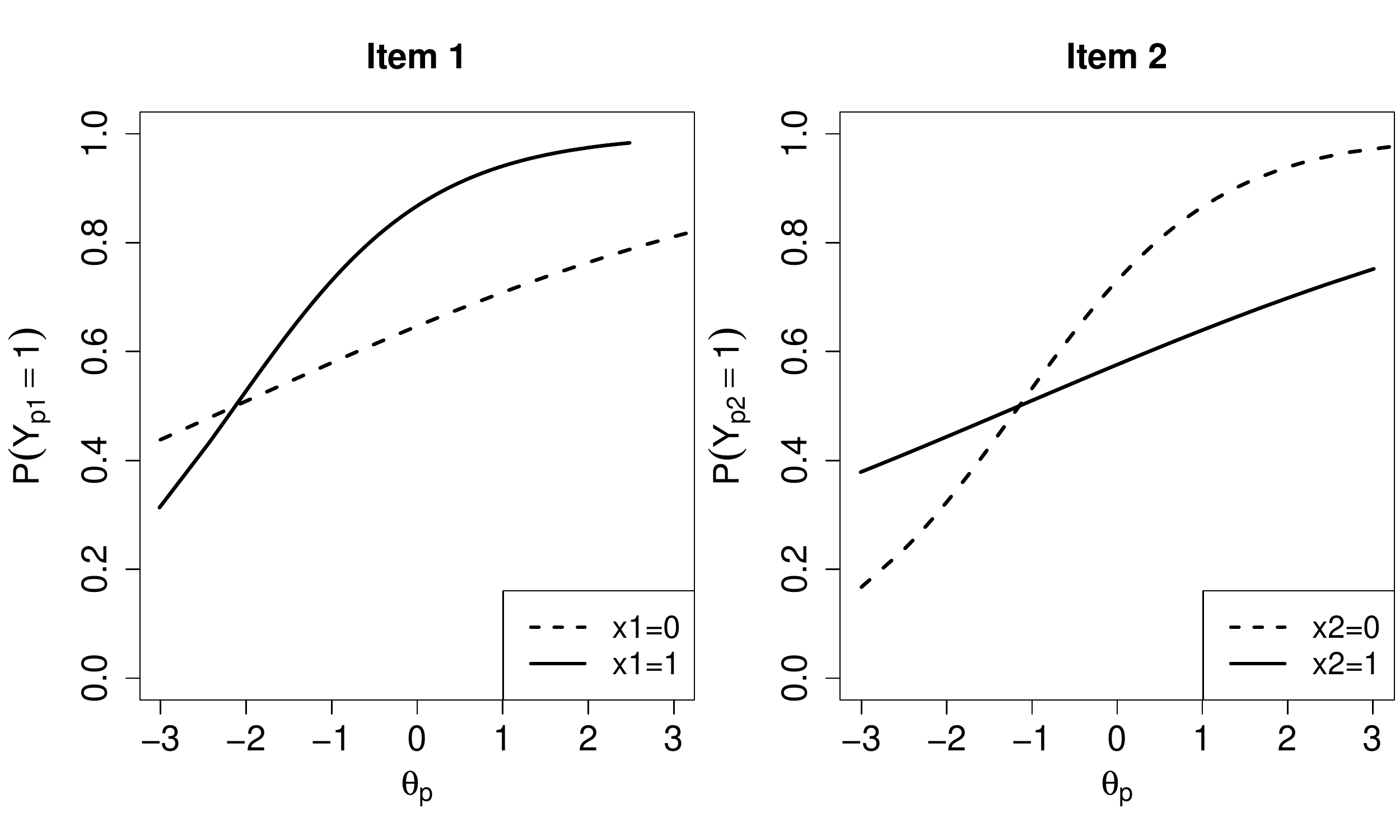}
\caption{Item Characteristic Curves of item 1 and item 2 for the the illustrative example with non-uniform DIF.}
\label{fig:example_nonuni_ICC}
\end{figure}

\subsection{Illustrative Example}

As in section \ref{sec:illustrative}  we consider data
$Y_{pi},\;p=1,\hdots,800,\;i=1,\hdots,20$, that are generated by a 2PL-model with DIF.
As before the item discrimination parameters $a_i$ are first drawn from a uniform distribution. However, in order to simulate non-uniform DIF we do not generate data from the 2PL-model but assume that the item discrimination parameters depend on covariates. The same strategy for generating non-uniform DIF was also used by \citet{rogers1993}, \citet{narayanan1996} or \citet{jodoin2001}.

Again, we consider 100 data sets with three covariates, two binary variables $x1,\,x2 \sim B(1,0.5)$ and one standard normal distributed variable $x3 \sim N(0,1)$. We simulate data where two of the 20 items show non-uniform DIF and two of the 20 items only show uniform DIF. The modified values of the discrimination and difficulty parameters are determined by step function given in Table \ref{tab:example_nonuni_sf}.
In item 1 and 3 DIF is induced by $x1$ and in item 2 and 4 DIF is induced by $x2$. Hence, in all four cases two groups have to be distinguished. The resulting ICC of the two items with non-uniform DIF (item 1 and 2) are given in Figure \ref{fig:example_nonuni_ICC} separately for the two groups. It can be seen from the curves that the item locations are equal for both groups but the item discriminations (as it was simulated) differ between the groups. When fitting \textit{IFT} the non-uniform DIF structure is detected correctly if there is one split in the slope component of the model of item 1 in $x1$ and item 2 in $x2$.

\subsubsection*{DIF}
\begin{figure}[!ht]
\centering
\includegraphics[width=0.35\textwidth]{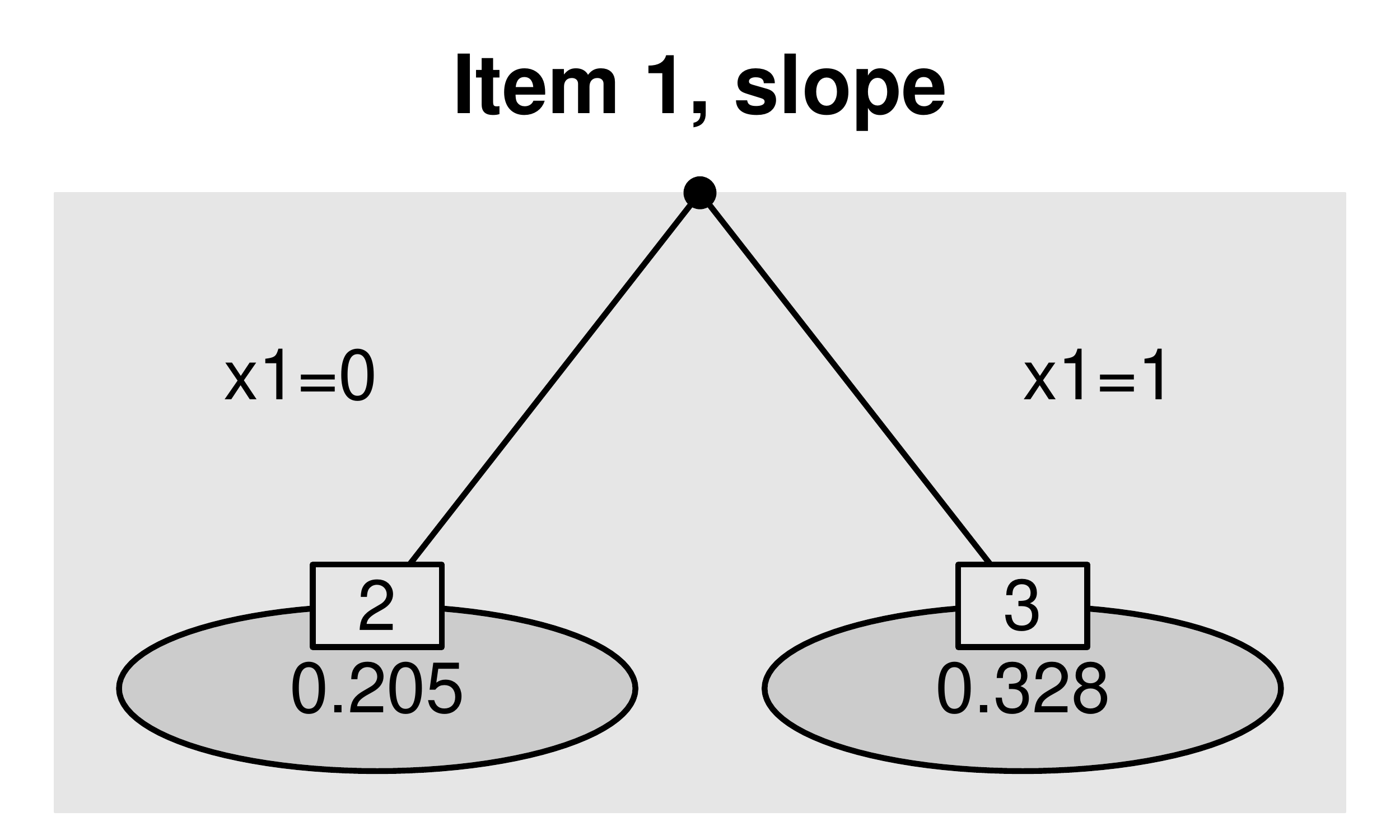}
\includegraphics[width=0.35\textwidth]{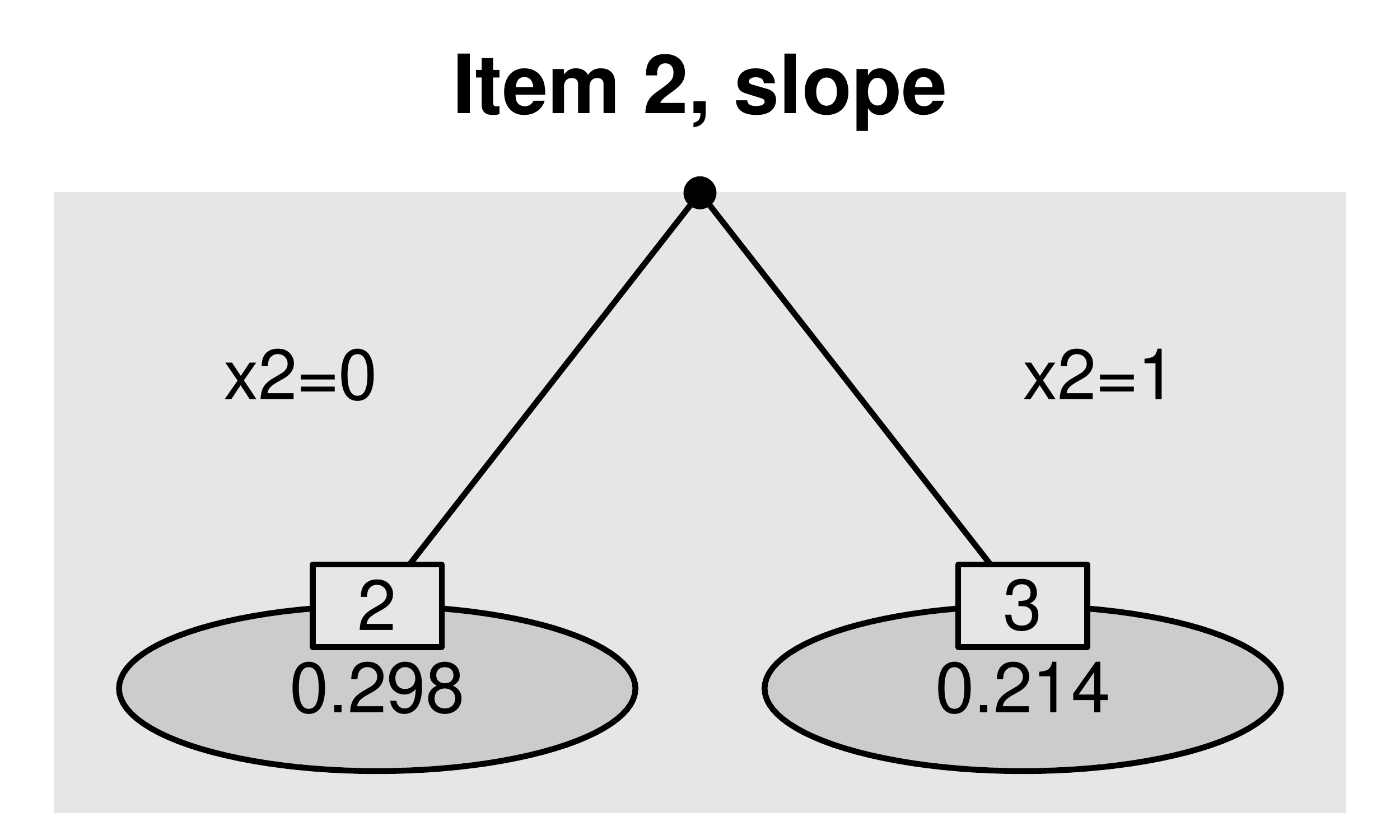}

\includegraphics[width=0.35\textwidth]{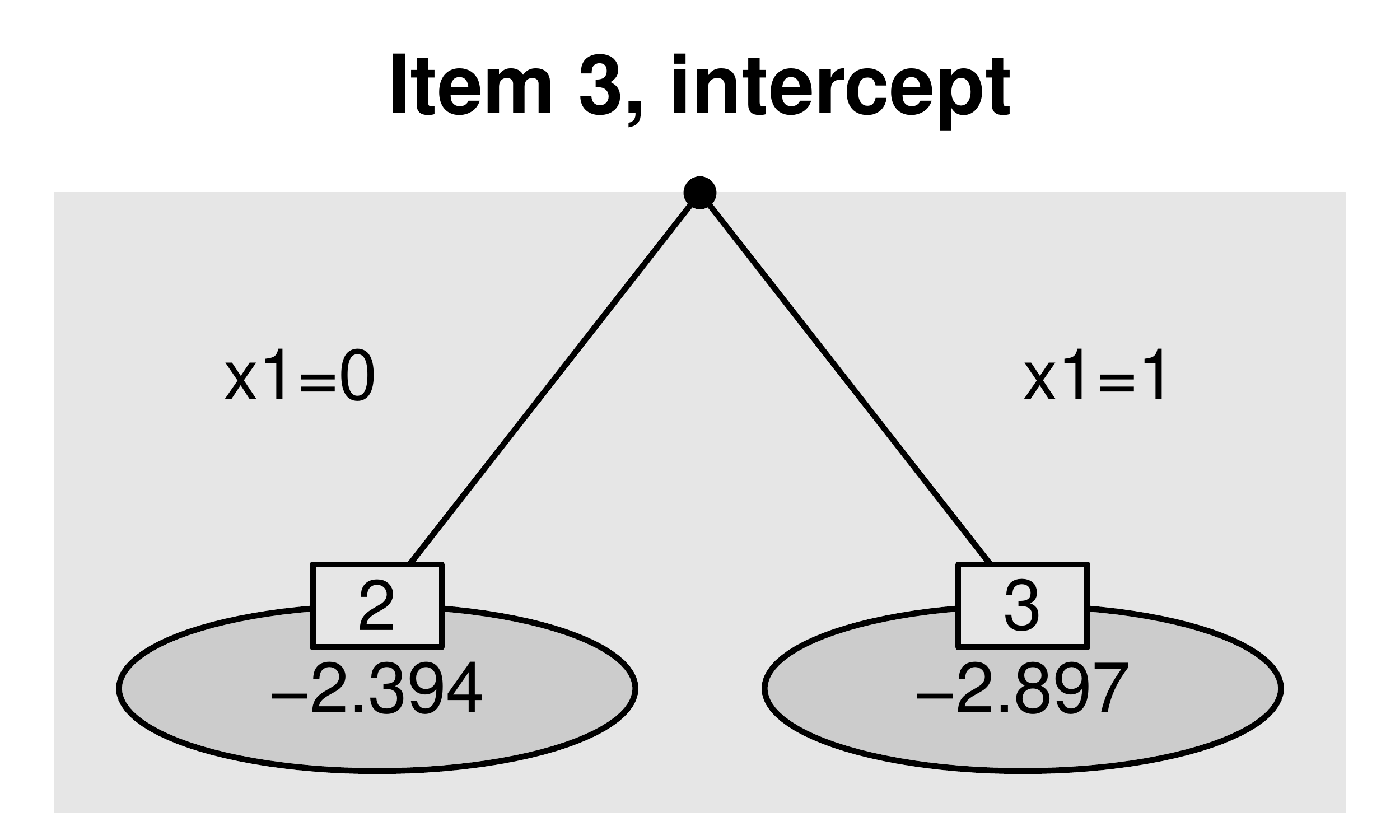}
\includegraphics[width=0.35\textwidth]{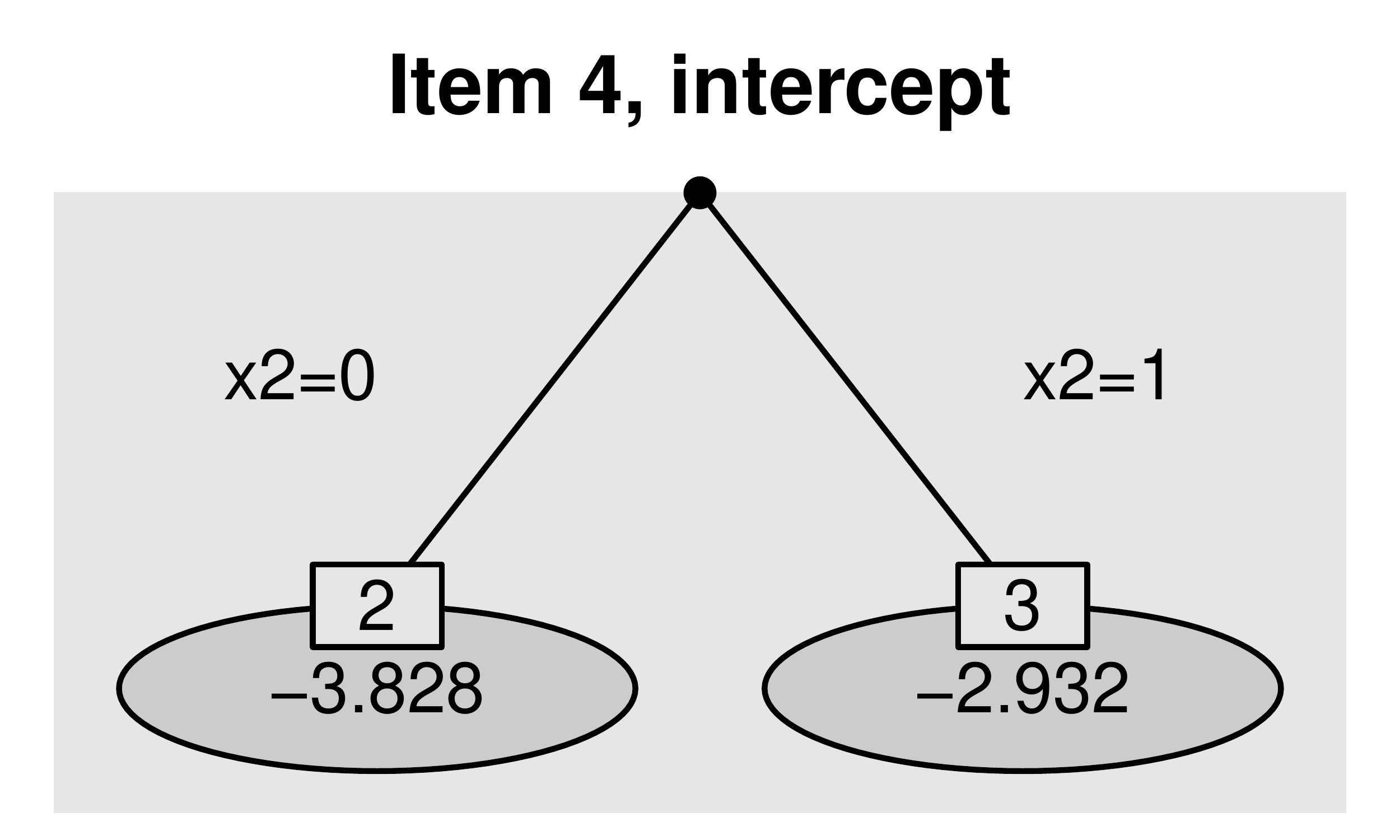}
\caption{Estimated trees for the illustrative example with non-uniform DIF, testing for both types of DIF. Estimated coefficients $\alpha_{i\ell}$ (upper) and $\gamma_{i\ell}$ (lower) are given in each leaf of the trees.}
\label{fig:example_DIF}
\end{figure}

Figure \ref{fig:example_DIF} shows one exemplary estimation result obtained by \textit{IFT} when testing for both types of DIF simultaneously. In this example items 1, 2, 3, and 4 are correctly identified as DIF items. All items are split once yielding trees with two terminal nodes, respectively. Items 1 and 2 (upper panel) are split with regard to the slopes indicating non-uniform DIF. In item 1 the (simulated) item discrimination is higher for $\{x1=1\}$, yielding a higher slope for the corresponding subgroup ($\hat{\alpha}_{1,x1=1}=0.328$). Whereas, in item 2 the item discrimination is larger for $\{x2=0\}$, which results in a larger slope for this subgroup ($\hat{\alpha}_{2,x2=0}=0.298$). In items 3 and 4 (lower panel) one split is performed with regard to the intercepts indicating uniform DIF. The  results are also in line with the true simulated effects. The model provides an identification of DIF items together with the responsible covariates and a classification by type of DIF.

\subsubsection*{Non-Uniform DIF}
\begin{figure}[!ht]
\centering
\includegraphics[width=0.35\textwidth]{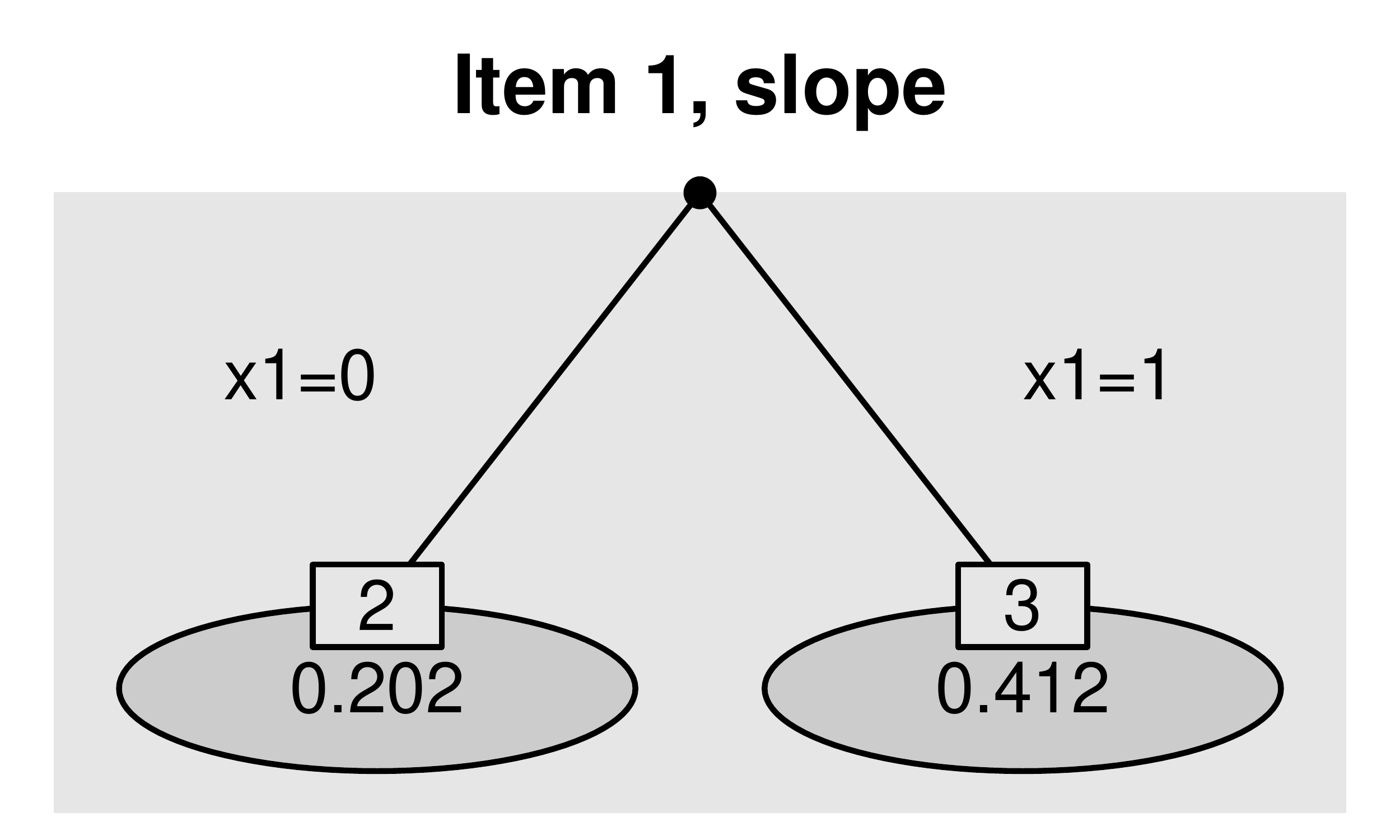}
\hspace{0.6cm}
\includegraphics[width=0.30\textwidth]{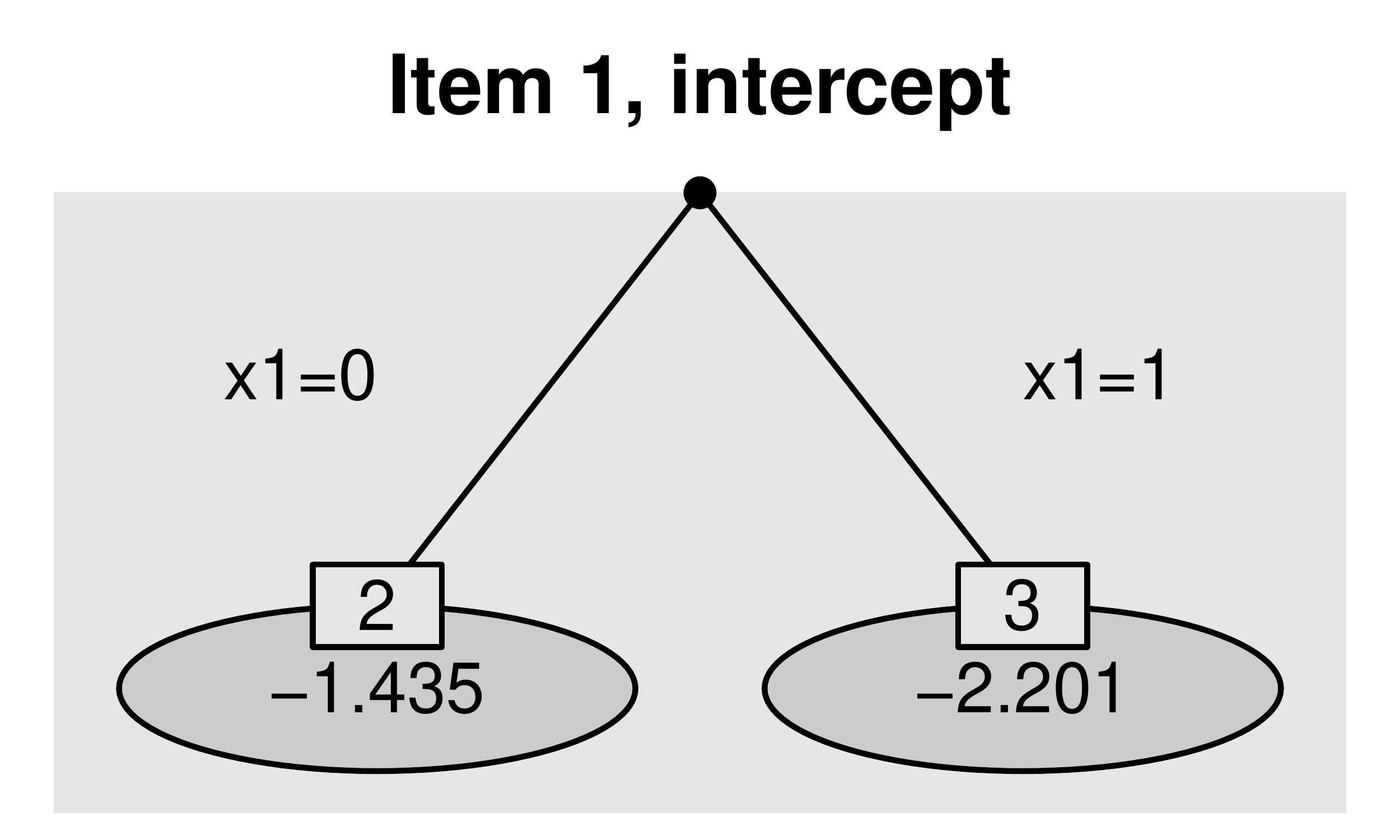}

\includegraphics[width=0.35\textwidth]{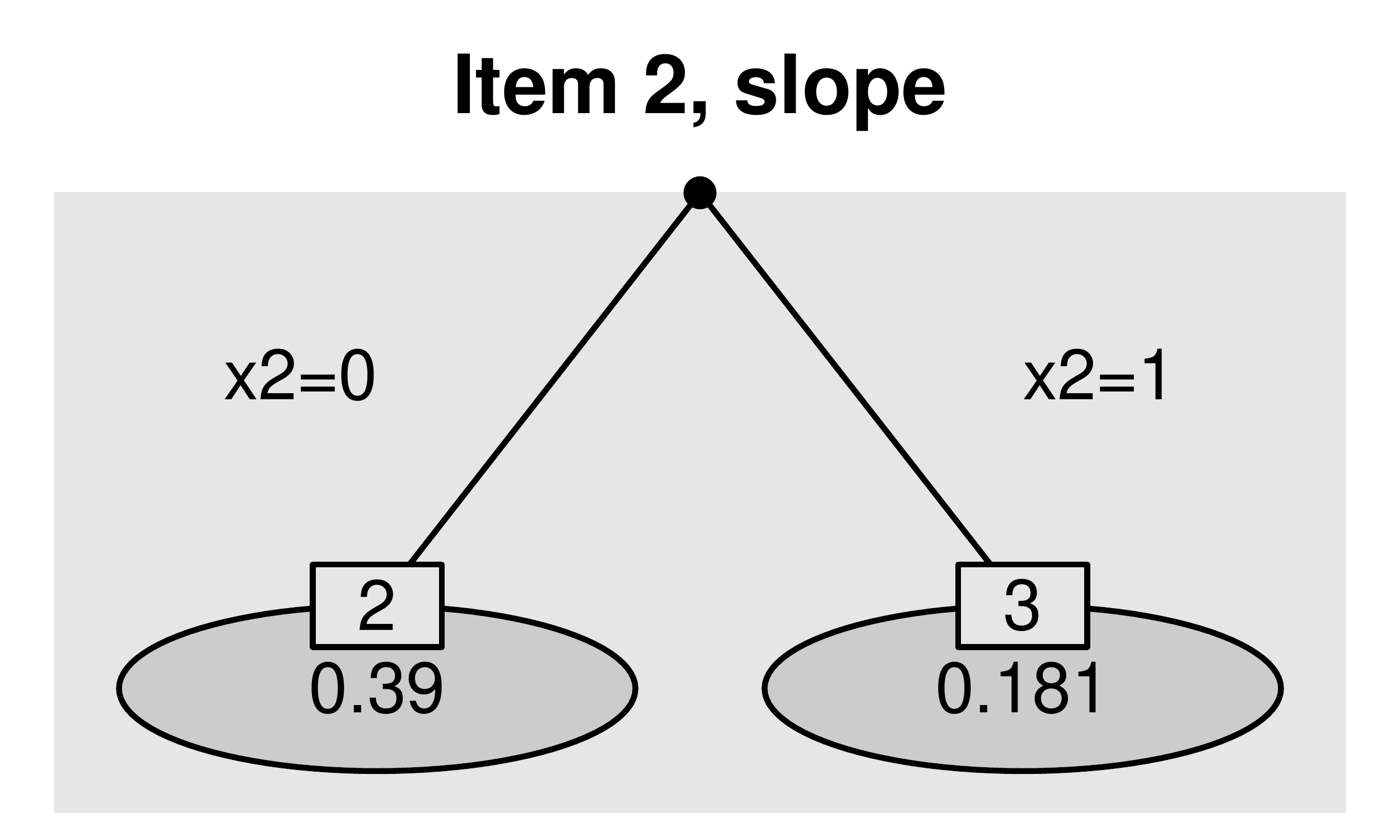}
\hspace{0.6cm}
\includegraphics[width=0.30\textwidth]{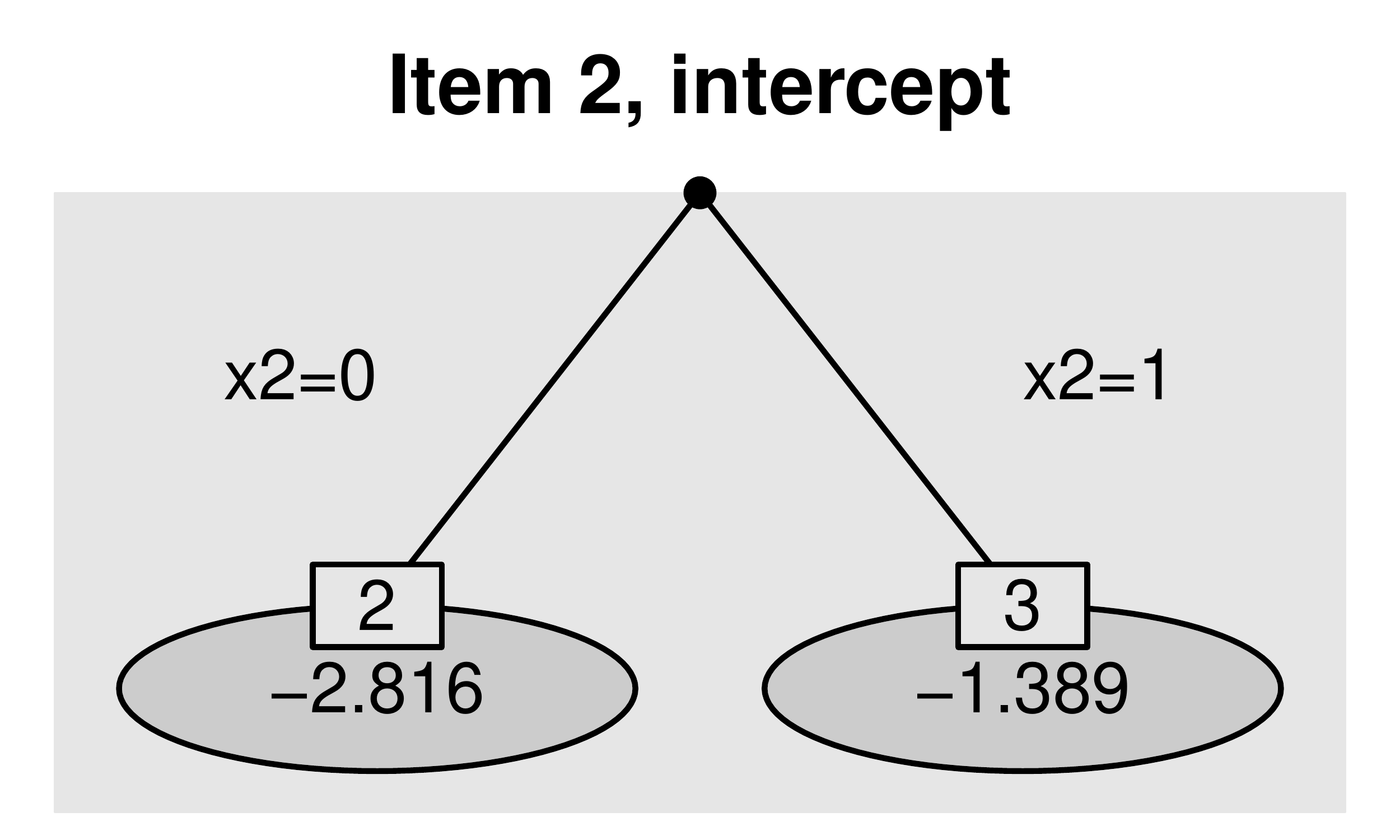}
\caption{Estimated trees for the illustrative example with non-uniform DIF, testing for non-uniform DIF. Estimated coefficients $\alpha_{i\ell}$ (left) and $\gamma_{i\ell}$ (right) are given in each leaf of the trees.}
\label{fig:example_UDIF}
\end{figure}

When using \textit{IFT}, which explicitly  tests for non-uniform DIF,  only items 1 and 2, that were simulated as non-uniform DIF items, are detected. The corresponding trees are given in Figure \ref{fig:example_UDIF}. The subgroup-specific slopes (left panel) are defined by the same splits as in  the \textit{DIF} framework considered previously. Due to the construction of the model the estimated coefficients $\alpha_{i1},\alpha_{i2},i=1,2$, however,  differ slightly. If splits are significant the same splits are performed in the intercepts yielding  trees with subgroup-specific intercepts. Since they are not of main interest they are displayed a little smaller (right panel of Figure \ref{fig:example_UDIF}).

\subsection{Simulations}

\begin{figure}[!ht]
\centering
\includegraphics[width=0.65\textwidth]{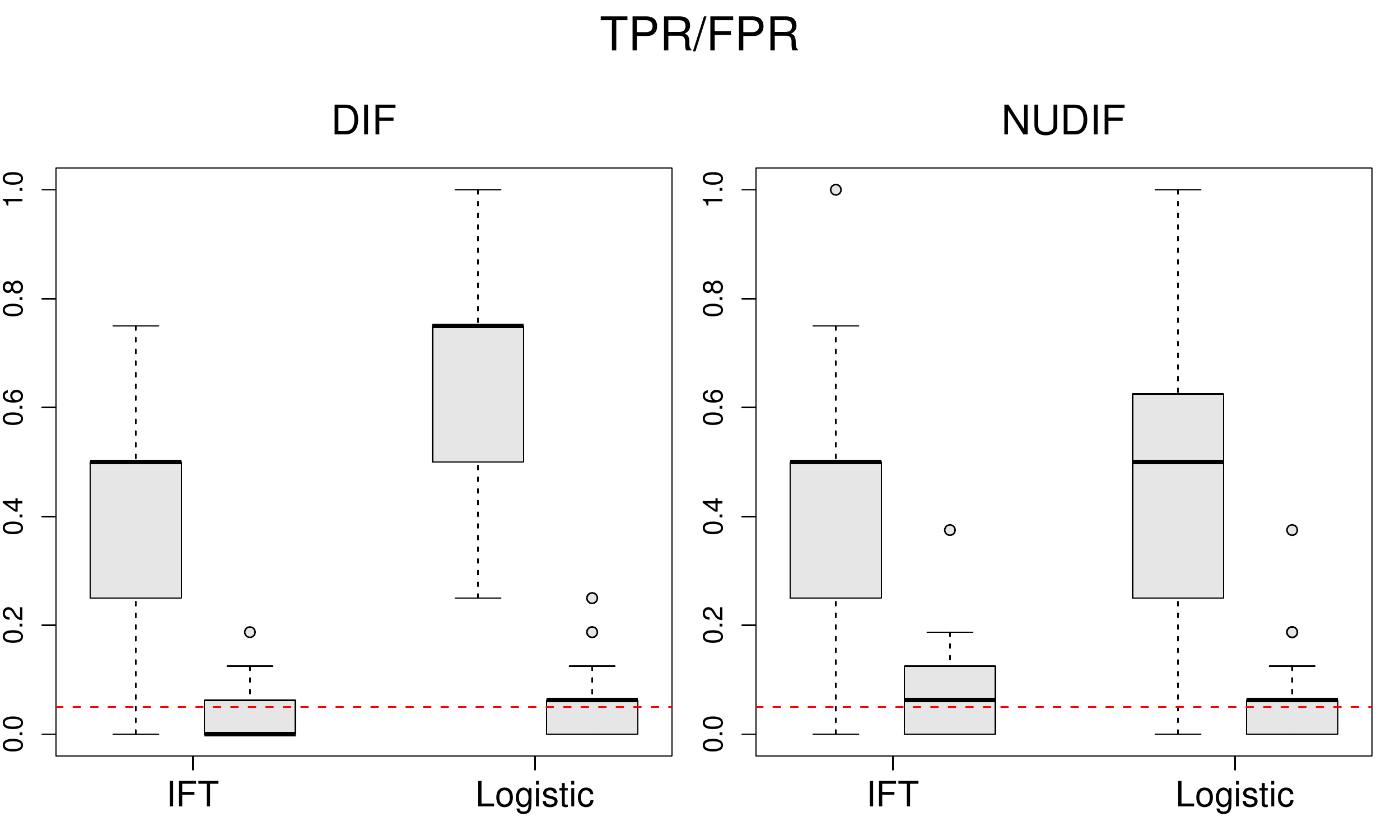}
\caption{Boxplots of TPR and FPR for the simulation with non-uniform DIF and one binary predictor ($P=800$, $I=20$, $20\%$ DIF), testing for both types of DIF (left) and testing for non-uniform DIF (right).}
\label{fig:sim1_nonuni_comp}
\end{figure}

In the following we briefly illustrate the properties of the models for the \textit{DIF} and \textit{NUDIF} framework by means of a small simulation. The structure of the simulated datasets we consider here is the same as in section \ref{sec:sim}. We limit the discussion to the comparison of two groups defined by one binary covariate $x\in\{0,1\}$. According to model \eqref{eq:2PL} non-uniform DIF is present if the item discriminations $a_i$ differ between the two groups. The difference in item discriminations is simulated by the equation $a_{i, \text{mod}}=a_i+ 0.6 \cdot I(x=0)$ for one half of DIF items and by the equation $a_{i, \text{mod}}=a_i+ 0.6 \cdot I(x=1)$ for the other half of DIF items.
Boxplots of true positive and false positive rates on the item level for the setting with $P=800$, $I=20$ and $20\%$ DIF obtained by \textit{IFT} (left of each panel) and the classical \textit{Logistic} model (right of each panel) are given in Figure \ref{fig:sim1_nonuni_comp}. The results when testing for both types of DIF are shown in the left panel and the results when testing for non-uniform DIF are shown in the right panel. Within the \textit{DIF} framework the classical \textit{Logistic} model outperforms the proposed tree based approach. The average hit rate in this setting is $0.66$ for \textit{Logistic} but only $0.43$ for \textit{IFT}. This  was to be expected because the test on the whole parameter vector $(\gamma_i,\alpha_i)$ obviously has a stronger power than the tests on single splits. However, in the \textit{NUDIF} framework the two methods almost yield the same results. The average hit rate for both models is $0.44$. Due to the construction of the models the main difference in the case of two groups is the use of different test statistics to obtain a decision. As we already illustrated for uniform DIF, our proposed tree based approach can also be used to detect non-uniform DIF without loss of efficiency.
The findings presented here can be confirmed by the results of all other settings considered in our simulation.

\section{Empirical Applications}

Finally we will illustrate and compare the proposed approaches on real data examples.

\begin{table}[!ht]
\caption{Summary statistics of the test score of the second module (items 21 to 40) of the I-S-T 2000 R and the two considered covariates.}
\begin{center}
\begin{tabularsmall}{lcccccc}
\toprule
Variable&\multicolumn{6}{c}{Summary statistics}\\
\midrule
&$x_{min}$&$x_{0.25}$&$x_{med}$&$\bar{x}$&$x_{0.75}$&$x_{max}$\\
Test score&6&12&14&13.87&16&19\\
Age&18&20&22&22.88&24&39\\
\\
Gender&\multicolumn{3}{c}{male: 97}&\multicolumn{3}{c}{female: 176}\\
\bottomrule
\end{tabularsmall}
\end{center}
\label{tab:app_ist_summary}
\end{table}

\begin{table}[!ht]
\caption{Comparison of detected DIF items of the I-S-T 2000 R using IFT and the extended Logistic approach for uniform and non-uniform DIF.}
\begin{center}
\begin{tabularsmall}{lcclcccc}
\toprule
&\multicolumn{4}{c}{\bf{Item focussed Trees}}&\multicolumn{3}{c}{\bf{Extended Logistic}}\\
\bf{Item}&UDIF&\multicolumn{2}{l}{DIF}&NUDIF&UDIF&DIF&NUDIF\\
\midrule
First&$\times$&$\times$&(u)&&$\times$&$\times$&\\
Second&$\times$&$\times$&(u)&&$\times$&$\times$&\\
Third&$\times$&$\times$&(non)&&$\times$&$\times$&\\
Fourth&&&&&$\times$&&\\
Fifth&&&&&$\times$&&\\
\bottomrule
\end{tabularsmall}
\end{center}
\label{tab:app_ist_comp}
\end{table}

\begin{figure}[!ht]
\begin{center}
\includegraphics[width=0.32\textwidth]{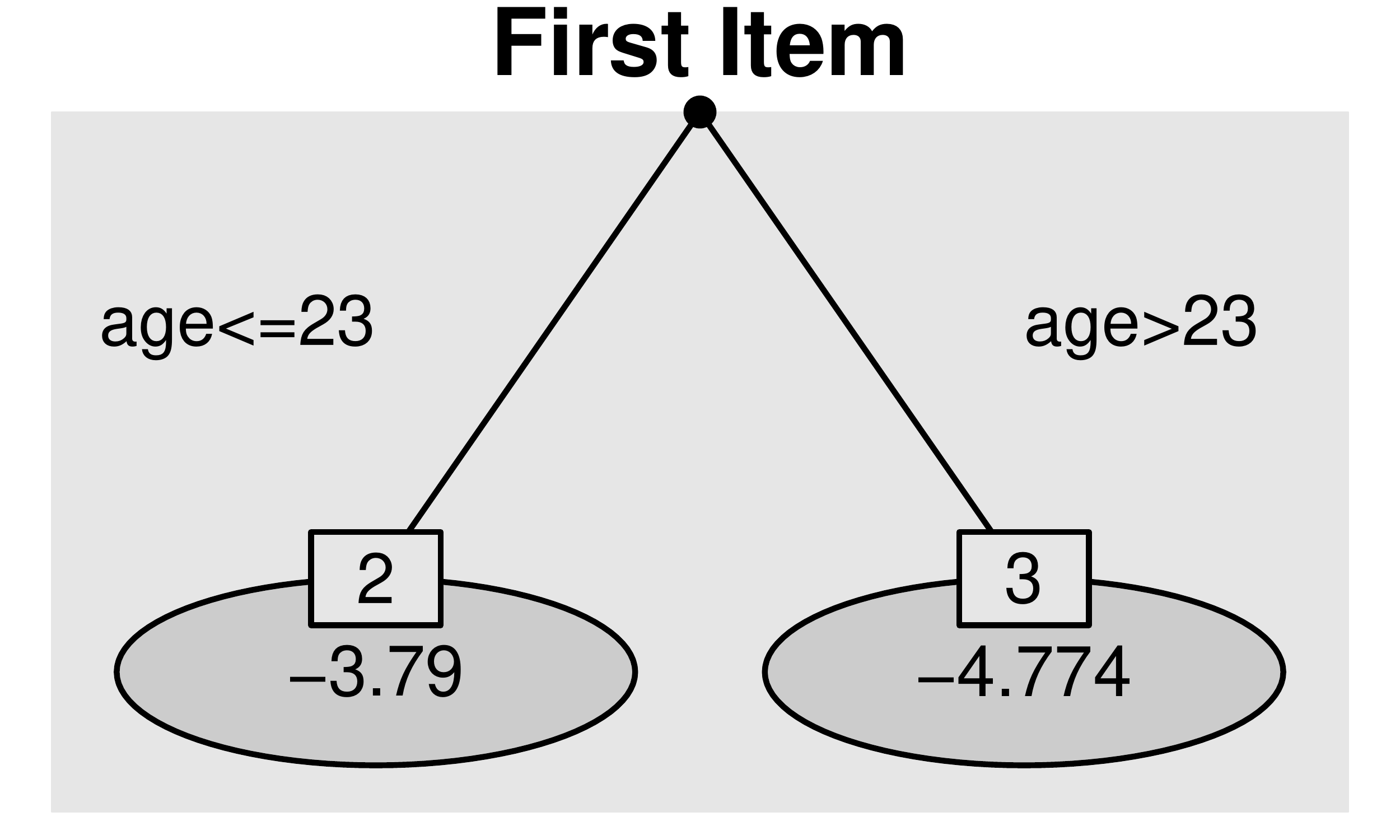}
\includegraphics[width=0.32\textwidth]{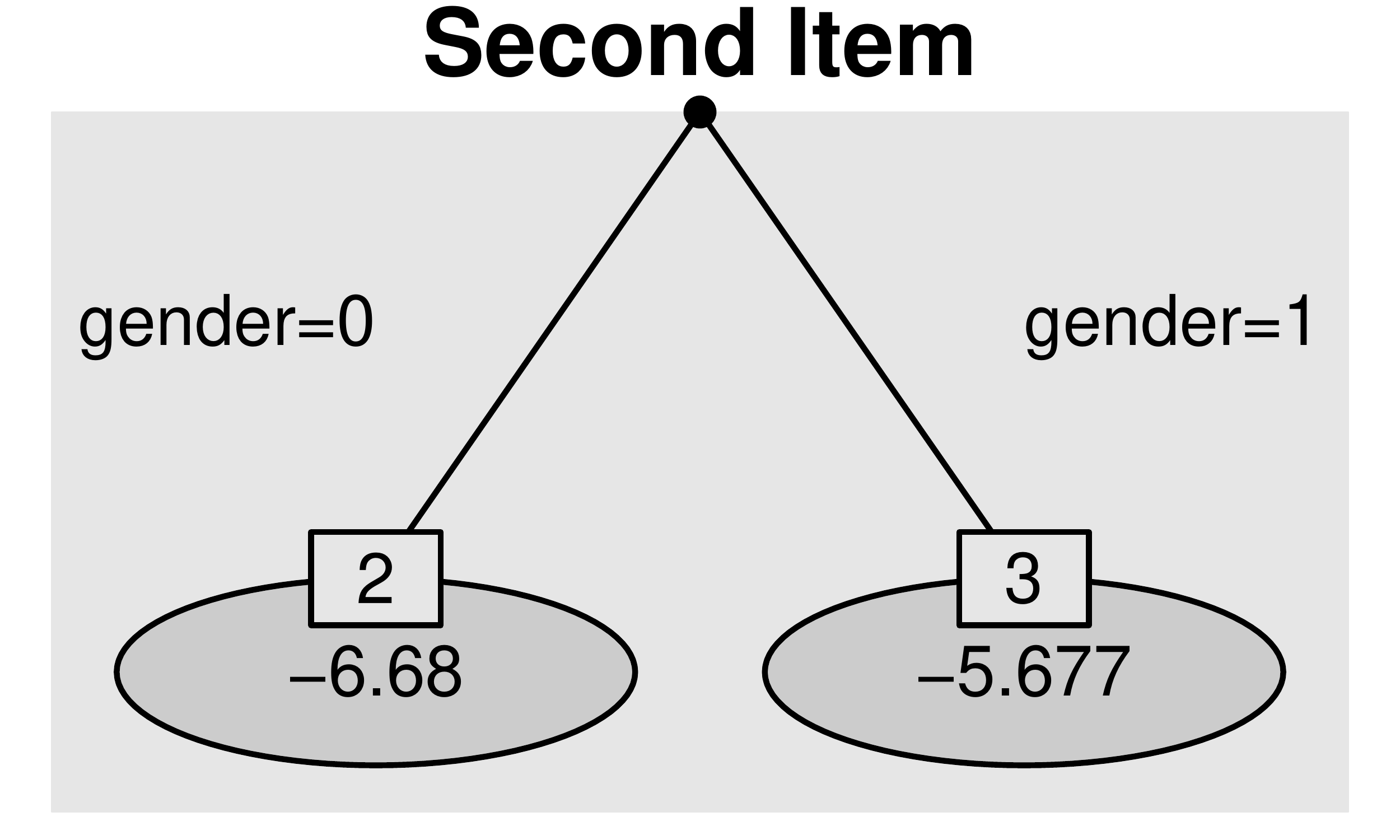}
\includegraphics[width=0.32\textwidth]{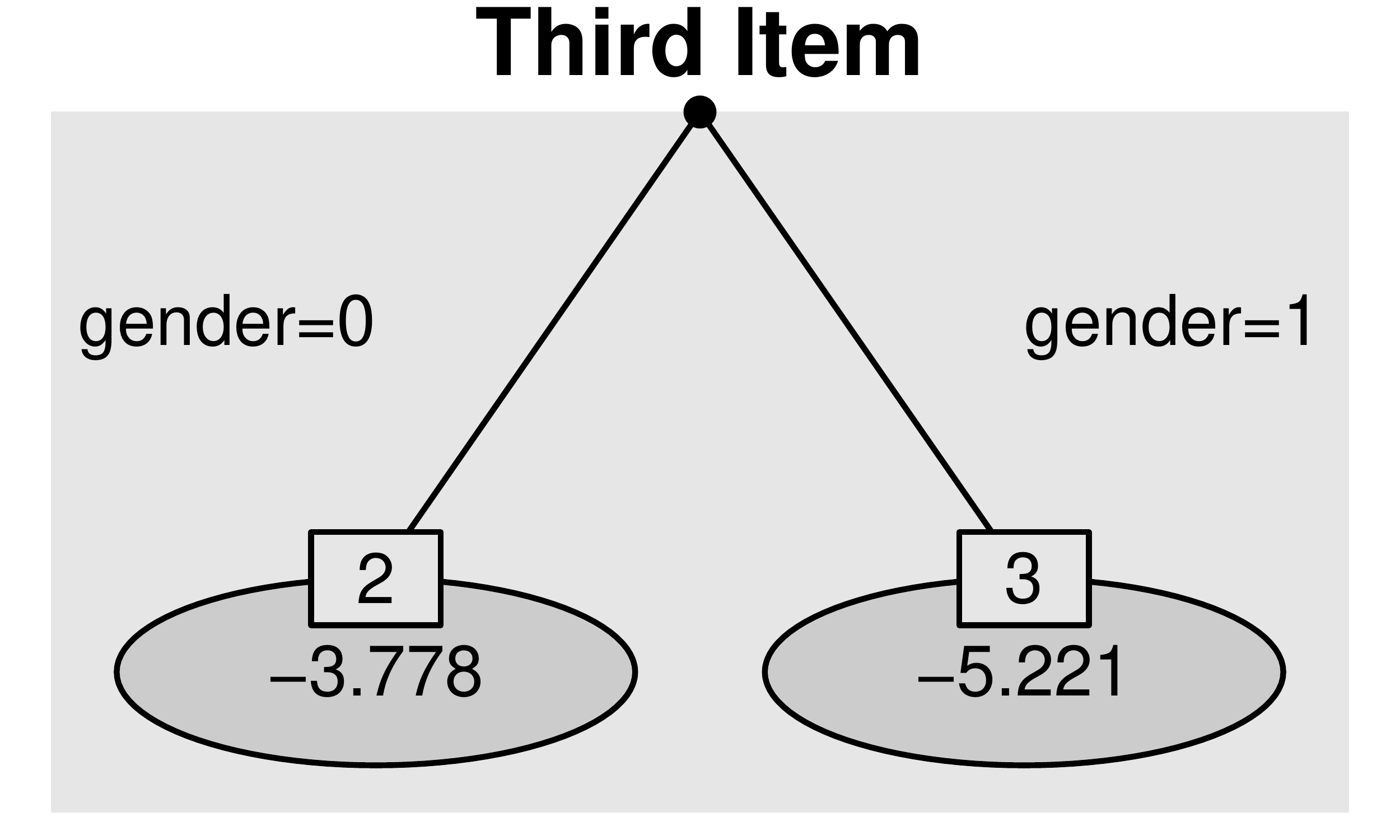}
\end{center}
\caption{Trees of the three detected DIF items of the second module of the I-S-T 2000 R using the model for uniform DIF. Estimated intercepts $\gamma_{il}$ are given in each leaf of the trees.}
\label{fig:tree_IST_283233}
\end{figure}

\begin{figure}[!ht]
\begin{center}
\includegraphics[width=0.4\textwidth]{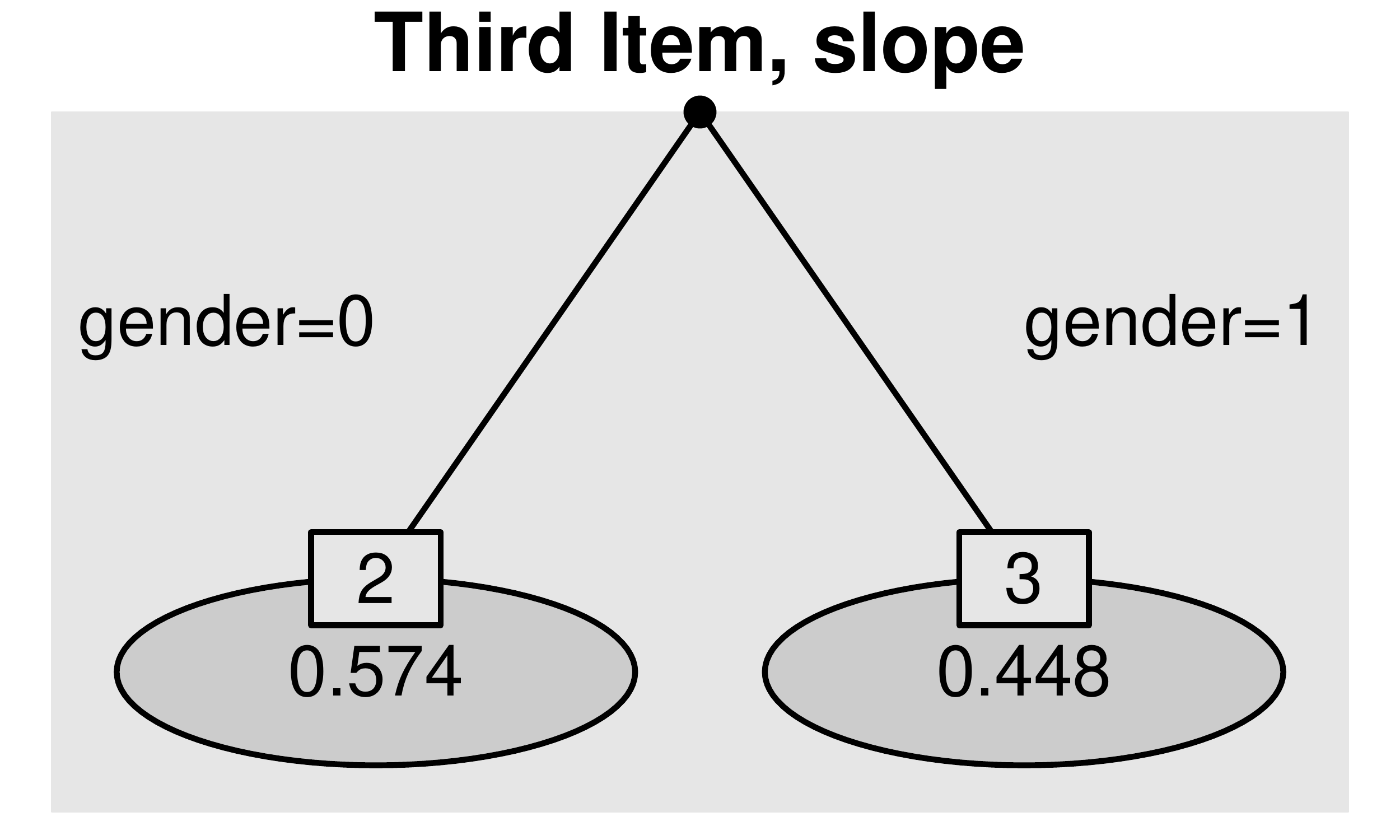}
\end{center}
\caption{Tree of the third detected DIF item of the second module of the I-S-T 2000 R using the model for both types of DIF. Estimated slopes $\alpha_{il}$ are given in each leaf of the trees.}
\label{fig:tree_IST_33non}
\end{figure}

\subsection{I-S-T 2000 R}
We use data from the Intelligence-Structure-Test 2000 R (I-S-T 2000 R; source of supply is Testzentrale G\"ottingen, Herbert-Quandt-Str. 4, 37081 G\"ottingen, Tel. (0049-551) 999-50-999, www.testzentrale.de). The test was developed by \citet{Amthauer2001, ist10english} and is a revised version of its predecessors I-S-T 70 \citep{ist70} and I-S-T 2000 \citep{Amthauer1999}.
The available study was conducted at the Phillips University in Marburg  \citep{Buhner2006}. There were 273 participants from 40 different subject areas. The second module of the test contains 20 items (items 21 to 40) in which  analogies play the major role. There are three predefined terms with a certain relation between the first two. This relationship needs to be recognized to find the fourth term. From five possible answers the respondent is asked to choose the term that  relates to the third term as the second term relates to the first term. One  example is
\begin{itemize}
\item[]{\bf dark:bright = wet:?}

a) rain $\,$ b) day $\,$ c) moist $\,$ d) wind $\,$ e) dry.
\end{itemize}
Therefore, one has to select that alternative that relates to wet as bright relates to dark.

For the investigation of DIF in these items we incorporate the covariates gender (male: 0, female: 1) and age. The summary statistics of the resulting test scores of items 21 to 40 and the two covariates are given in Table \ref{tab:app_ist_summary}.

When using \textit{IFT} for uniform DIF  3 out of 20 items showed DIF. The algorithm performs only three splits before stopping and, therefore, each item is split only once. All permutation tests were based on 1000 permutations at local significance level $0.05/2$.

The estimated trees for three items detected as DIF items are given in Figure \ref{fig:tree_IST_283233}. It is seen that both covariates gender and age seem to induce DIF because both are  used for splitting at least once. The second and third item show DIF induced by gender, whereas the first item shows DIF induced by age. According to the estimated coefficients the second item is easier for females (gender=1), the third item is easier for males (gender=0) and the first item is easier for all students who are rather young (age$\leq$23).

An overview of the detected DIF items obtained by the six strategies discussed in this article is given in Table \ref{tab:app_ist_comp}.
When using \textit{IFT} which tests for both types of DIF, one obtains very similar results. As in the \textit{UDIF} framework the first, second and third item are also identified as DIF items  with the same variables that induce DIF. The estimated models for the first and second item are even identical. A difference occurs for the third item, where the split in gender is not performed in the intercept but in the slope component. The model gives the estimated intercept $\beta_{0,Third}=-4.993$. The resulting tree of slopes $\alpha_{il}$ is given in Figure \ref{fig:tree_IST_33non}. The estimated coefficients again mean that the item favours males (gender=0) but the difference slightly increases for participants with a higher test score.
Interestingly, the splits in the intercept (\textit{UDIF}, Figure \ref{fig:tree_IST_283233}) and in the slope (\textit{DIF}, Figure \ref{fig:tree_IST_33non}) result in very similar estimated probabilities. As a consequence it is not surprising that the third item is not detected by the model within the \textit{NUDIF} framework.

The evaluation of the data set by the extended \textit{Logistic} model \eqref{eq:linear_uniform} for uniform DIF yields five DIF items (fourth column in Table \ref{tab:app_ist_comp}). Based on the results in the simulations, it seems that the fourth and fifth item might be falsely identified as items with uniform DIF. Concerning the identification of items, the results within the \textit{DIF} and \textit{NUDIF} framework are equal to those of \textit{IFT}. However, when testing non-uniform DIF for the third item one obtains the $p$-value $0.052$ indicating a significant effect.
It should again be mentioned, that despite of the similar results, the extended \textit{Logistic} approach does not provide any information about the variables that are responsible for DIF.

It is noteworthy that in summary the test seems not to be  strongly affected by DIF. From the 20 items that use analogies only three are suspect of DIF and the effects are not overly strong. This was to be expected of a carefully designed test.

\begin{table}[!ht]
\caption{Summary statistics of the test score of the 25 multiple-choice items from subject area science of the CTB data and the three considered covariates.}
\begin{center}
\begin{tabularsmall}{lcccccc}
\toprule
Variable&\multicolumn{6}{c}{Summary statistics}\\
\midrule
&$x_{min}$&$x_{0.25}$&$x_{med}$&$\bar{x}$&$x_{0.75}$&$x_{max}$\\
Test score&7&14&16&16.01&18&23\\
Size&100&500&900&868.3&1300&1600\\
\\
Type&\multicolumn{2}{c}{catholic: 105}&\multicolumn{2}{c}{private: 84}&\multicolumn{2}{c}{public: 1311}\\
Gender&\multicolumn{3}{c}{male: 761}&\multicolumn{3}{c}{female: 739}\\
\bottomrule
\end{tabularsmall}
\end{center}
\label{tab:app_ctb_summary}
\end{table}

\begin{figure}[!ht]
\centering
\includegraphics[width=0.6\textwidth]{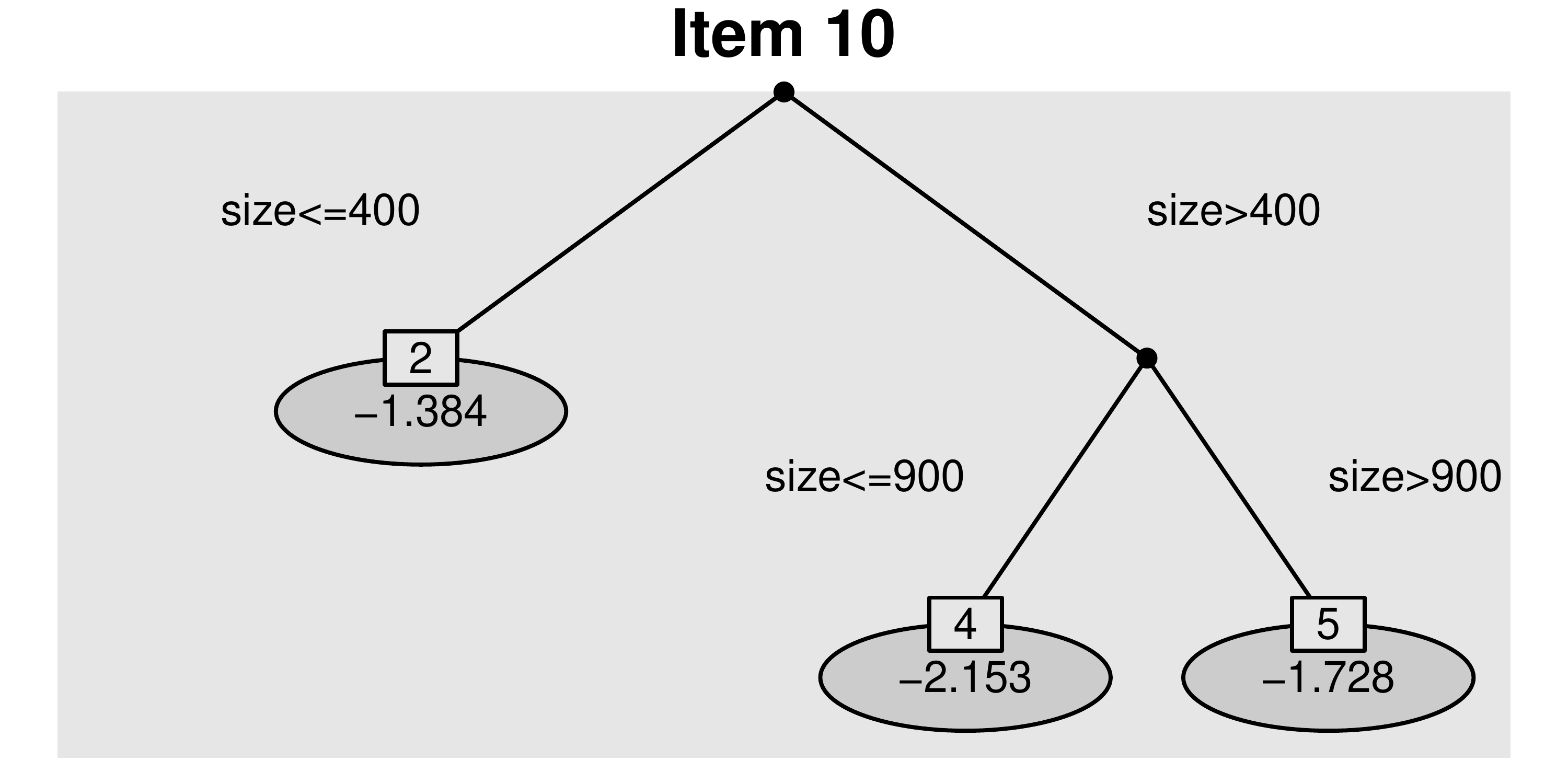}
\includegraphics[width=0.32\textwidth]{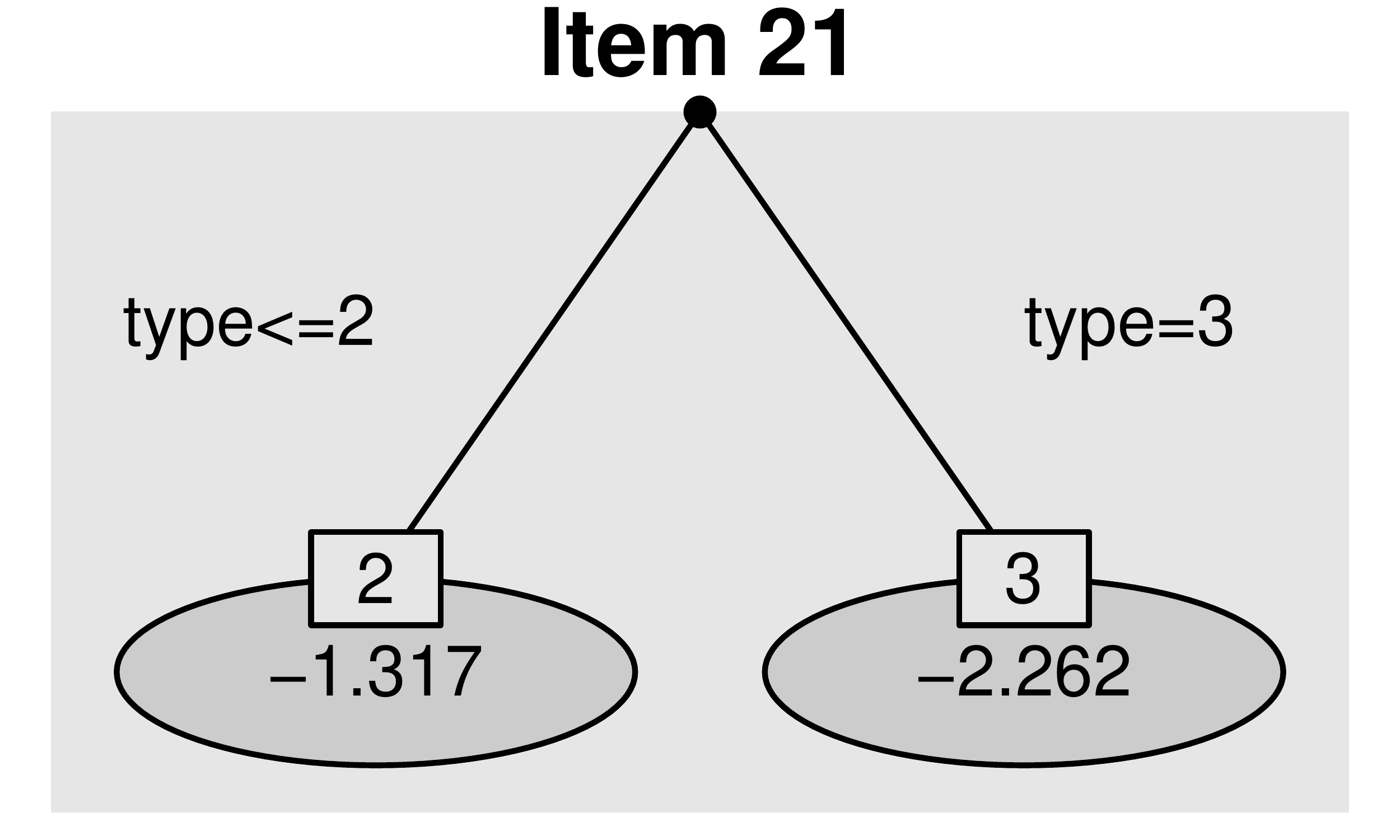}
\includegraphics[width=0.8\textwidth]{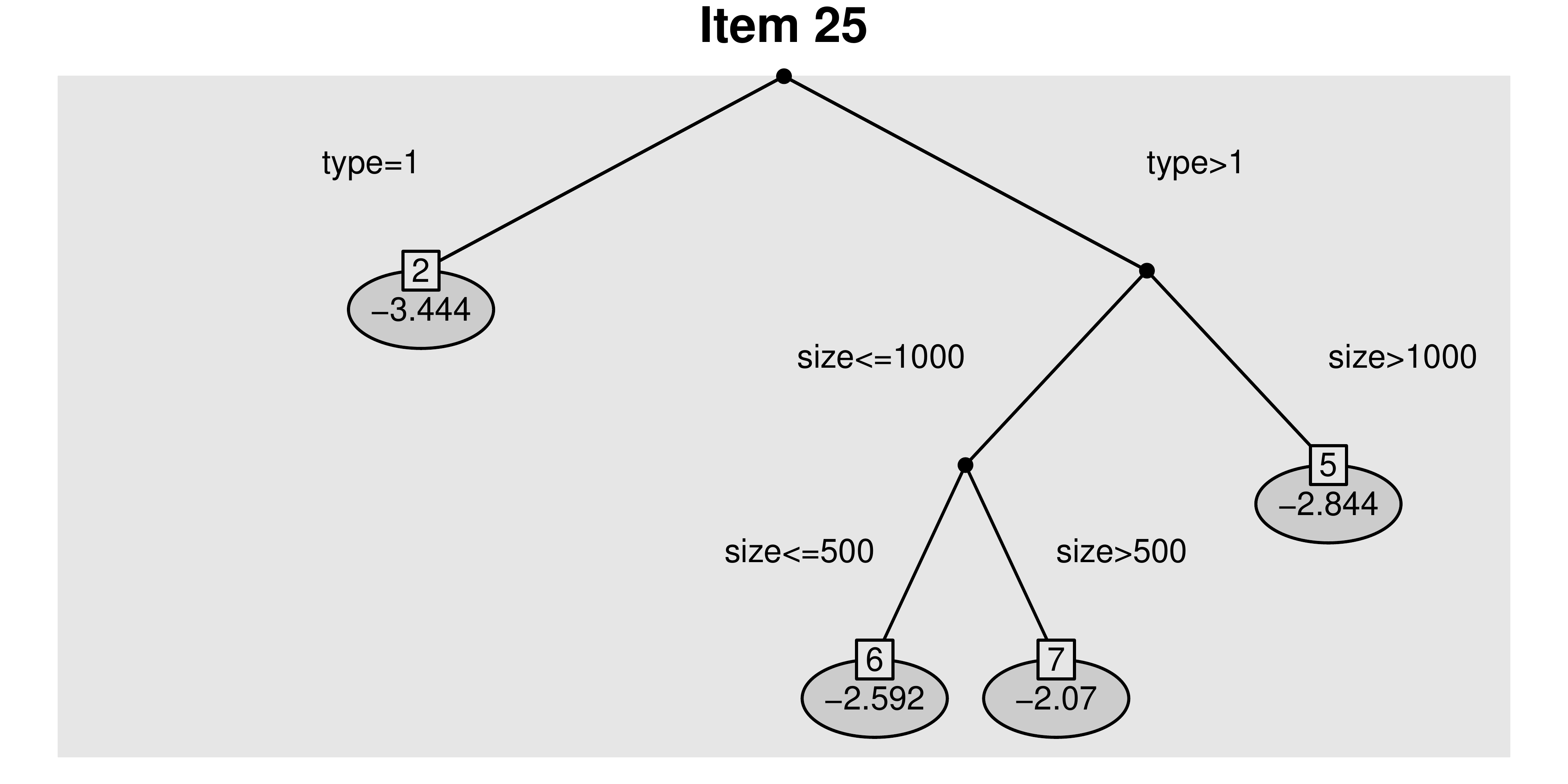}
\caption{Trees of items 10, 21 and 25 of the CTB data using the model for uniform DIF. Estimated intercepts $\gamma_{il}$ are given in each leaf of the trees.}
\label{fig:tree_CTB_102125}
\end{figure}

\begin{table}[!ht]
\caption{Comparison of detected DIF items of the CTB data using IFT and the extended Logistic approach for uniform and non-uniform DIF.}
\begin{center}
\begin{tabularsmall}{lcclcccc}
\toprule
&\multicolumn{4}{c}{\bf{Item focussed Trees}}&\multicolumn{3}{c}{\bf{Extended Logistic}}\\
\bf{Item}&UDIF&\multicolumn{2}{l}{DIF}&NUDIF&UDIF&DIF&NUDIF\\
\midrule
21&$\times$&$\times$&(non)&$\times$&$\times$&$\times$&$\times$\\
3&$\times$&$\times$&(u)&&$\times$&$\times$&\\
4&$\times$&$\times$&(u)&&$\times$&$\times$&\\
8&$\times$&$\times$&(u)&&$\times$&$\times$&\\
9&$\times$&$\times$&(u)&&$\times$&$\times$&\\
14&$\times$&$\times$&(non)&&$\times$&$\times$&\\
16&$\times$&$\times$&(non)&&$\times$&$\times$&\\
25&$\times$&$\times$&(u)&&$\times$&$\times$&\\
11&&&&&$\times$&$\times$&$\times$\\
13&&&&$\times$&&$\times$&$\times$\\
19&$\times$&$\times$&(u)&&$\times$\\
5&$\times$&$\times$&(u)&&&\\
10&$\times$&$\times$&(u)&&&\\
24&&&&&$\times$&$\times$&\\
1&&&&&&&$\times$\\
6&$\times$&&&&&&\\
15&$\times$&&&&&&\\
17&$\times$&&&&&&\\
\bottomrule
\end{tabularsmall}
\end{center}
\label{tab:app_ctb_comp}
\end{table}

\subsection{CTB Science data}

In a second application we consider a data set from CTB-McGraw Hill. For a description of the original data, see also \citet{de2004explanatory}. The data includes the results of 1500 grade 8 students from 35 schools. The students had to respond to 76 items, measuring different objectives and subskills related to mathematics and science. In our investigation we restrict to the 25 multiple-choice items from subject area science.

To test for DIF in these items we incorporate the three covariates gender (male: 0, female: 1), type of the school (1: catholic, 2: private, 3: public) and size of the school (number of students in hundreds). The summary statistics of the test scores for the 25 items and the three covariates are given in Table \ref{tab:app_ctb_summary}.

When fitting \textit{IFT} for uniform DIF 14 of 25 items are identified as DIF items. Altogether the algorithm performs 27 splits until further splits are no longer significant. With three covariates, each permutation test is performed at local significance level $0.05/3$. The $p$-value in the $28$th iteration was $0.02$ and thus not significant on level $0.01\overline{6}$. All splits refer to covariates type and size, whereas no significant splits were found for variable gender. There does not seem to be any difference between males and females.

The trees for three selected items are given in Figure \ref{fig:tree_CTB_102125}.
In item 10 DIF is induced by size and one has to distinguish between three subgroups. The item is easiest for students in small schools (size$\leq$400) but most difficult for students in medium-sized schools (400$<$size$\leq$900). Item 21 is easier for students in a catholic or private school (type$\leq$2) compared to students in public schools (type$=$3). An interesting partition is received for item 25. For all students in a catholic school (type$=$1) the question is very difficult. By contrast the question is easier for all students in a private or public school (type$>$1), in particular for those in medium-sized schools (500$<$size$\leq$1000).

An overview of the detected DIF items by the six evaluated models is given in Table \ref{tab:app_ctb_comp}. It shows only items that were found to be DIF items by at least one of the models. Within the \textit{DIF} framework (second column) eleven DIF items are identified. These are the same items as with the restricted model for uniform DIF discussed above, but without item 6, 15 and 17. Unlike above, there are three items that are classified as non-uniform DIF items by the more general model. Here, for example in item 21 the split regarding the type of school is not performed in the intercept but in the slope component.
According to the model testing for non-uniform DIF (third column) the two items 13 and 21 carry non-uniform DIF. In contrast to item 13, item 21 is also detected within the \textit{UDIF} and \textit{DIF} framework.

The comparison to the extended \textit{Logistic} approach shows a strong overlap. Within the \textit{UDIF} framework (first and fourth column) there is a agreement in nine items. In the \textit{DIF} framework this is the case for eight items. However it should again be mentioned, that the extended \textit{Logistic} approach within the \textit{DIF} framework does not distinguish between uniform and non-uniform DIF. When testing for non-uniform DIF (sixth column) one obtains four significant results. In contrast to items 1 and 11, items 13 and 21 are also found by \textit{IFT}. In total item 21 is the only item that shows DIF according to all six models and four items are only identified as DIF items by one of the six models.

\section{Concluding Remarks}

The proposed recursive partitioning approach, in short IFT, is an extension of the basic logistic regression model for the detection of uniform and non-uniform DIF. In contrast to the classical approach, IFT allows to incorporate several covariates on different scales, including ordinal and continuous covariates, that potentially induce DIF. The method leads to simultaneous selection of items and (interactions of) variables that cause DIF. The result typically is a small tree for each DIF item and therefore the DIF structure is easy accessible.

The results of the simulations including uniform as well as non-uniform DIF show that IFT has the same performance than the classical approach in the simple case of two groups but also works quite well in more complex settings with various covariates.
The applications demonstrate the flexibility and interpretability of IFT, also compared to the extended Logistic model that tests DIF by a vector of covariates. In particular, within the framework that tests for both types of DIF the obtained trees show which type of DIF is present.

The results shown in the paper were obtained by an R program, that is available by the authors and will soon be available in an extended version of the R add-on package \texttt{DIFtree} \citep{DIFtree2015} on CRAN.

\bibliography{literatur}

\end{document}